\documentclass[11pt]{amsart}

\usepackage{amssymb}
\usepackage{amsmath}
\usepackage{amsthm}
\usepackage{amsfonts}
\usepackage{titlesec}
\usepackage{mathabx}
\usepackage{xcolor}
\usepackage{graphicx}
\usepackage{subcaption}
\usepackage{float}
\usepackage{multirow}
\usepackage{todonotes}
\usepackage{algorithmic}
\usepackage{algorithm}
\usepackage{tikz}
\usepackage{url}
\usepackage[nocompress]{cite}

\newtheorem{thm}{Theorem}[section]

\newtheorem{lem}[thm]{Lemma}
\newtheorem{prop}[thm]{Proposition}

\newtheorem{rem}[thm]{Remark}

\newtheorem{ass}[thm]{Assumption}

\titleformat{\section}{\normalfont\bfseries\centering}{\thesection.}{.25em}{}
\titleformat{\subsection}{\normalfont\bfseries}{\thesubsection.}{.25em}{}
\titlespacing{\section}{0pt}{*5}{*1.5}
\titlespacing{\subsection}{0pt}{*3}{*0.5}
\numberwithin{equation}{section}
\allowdisplaybreaks
\newcommand{\R}{\ensuremath{\mathbb R}}    %
\newcommand{\N}{\ensuremath{\mathbb N}}    %

\newcommand{\V}{\ensuremath{\mathbb V}}  
\newcommand{\X}{\ensuremath{\mathbb X}}  
\newcommand{\bbP}{\ensuremath{\mathbb P}}
\newcommand{\calI}{\ensuremath{\mathcal I}}
\newcommand{\calN}{\ensuremath{\mathcal N}}    %
\newcommand{\calL}{\ensuremath{\mathcal L}}    %

\newcommand{\bbV}{\ensuremath{\mathbb V}}    %
\newcommand{\spn}{\ensuremath{\operatorname{span}}}
\newcommand{\mas}[2]{\left[\begin{array}{#1}#2\end{array}\right]}

\newcommand{\infn}[1]{\ensuremath{\|#1\|_{{\infty}}  }}
\newcommand{\romd}[1]{\ensuremath{\mathrm{d}{#1}}}
\usepackage{hyperref}
\newcommand\blfootnote[1]{%
	\begin{NoHyper}%
		\renewcommand\thefootnote{}\footnote{#1}%
		\addtocounter{footnote}{-1}%
	\end{NoHyper}%
}

\newcommand{\nametbdd}[1]{\textcolor{black}{#1}}
\newif\ifshowcommentyan 
\showcommentyantrue  %

\newif\ifshowcommentMS
\showcommentMStrue      %

\begin{document}
\title[]{Modularized data-driven approximation of the Koopman operator and generator$^*$} %
\blfootnote{$^*$Y.~Guo is funded by the European Social Fund Plus (ESF Plus) and the Free State of Saxony through the project HZwo:StabiGrid. K.~Worthmann gratefully acknowledges support from the Carl Zeiss Foundation (VerneDCt -- Project No.\ 2011640173).}

 \author[Y.\ Guo]{Yang Guo}
 \address{{\bf Y.~Guo:} Automatic Control and System Dynamics, Faculty of Electrical Engineering and Information Technology, Chemnitz University of Technology}
  
 \author[M.\ Schaller]{Manuel Schaller}
 \address{{\bf M.~Schaller:} Institute of Mathematics, Technische Universit\"at Ilmenau and Faculty of Mathematics, Chemnitz University of Technology}
 
  \author[K.\ Worthmann]{Karl Worthmann}
 \address{{\bf K.~Worthmann:} Optimization-based Control Group, Institute of Mathematics, Technische Universit\"at Ilmenau}

 \author[S.\ Streif]{Stefan Streif}
 \address{{\bf S.~Streif:} Automatic Control and System Dynamics, Faculty of Electrical Engineering and Information Technology, Chemnitz University of Technology}

\begin{abstract}
Extended Dynamic Mode Decomposition (EDMD) is a widely-used data-driven approach to learn an approximation of the Koopman operator. Consequently, it provides a powerful tool for data-driven analysis, prediction, and control of nonlinear dynamical (control) systems. 
In this work, we propose a novel modularized EDMD scheme tailored to interconnected systems. 
To this end, we utilize the structure of the Koopman generator that allows to learn the dynamics of subsystems individually  and thus alleviates the curse of dimensionality by considering observable functions on smaller state spaces. Moreover, our approach canonically enables transfer learning if a system encompasses multiple copies of a model as well as efficient adaption to topology changes without retraining. We provide finite-data bounds on the estimation error using tools from graph theory. The efficacy of the method is illustrated by means of various numerical examples. 
\end{abstract}

\maketitle


\section{Introduction}
\noindent The Koopman operator~\cite{koopman1931hamiltonian} provides a route to view the flow of a nonlinear dynamical system in an infinite dimensional space through the lens of observable functions. As a consequence of this lifting, the action of the Koopman operator is linear. When restricted to finite dictionaries of observables, i.e.\  a finite-dimensional subspace of functions, one may approximate the Koopman operator by means of measurement data only, e.g., by (Extended) Dynamic Mode Decomposition~((E)DMD; \cite{williams2015data,schmid2010dynamic}). Due to its accessibility through data-driven methods, the Koopman approach has attained a lot of attention in recent years, cf.~\cite{mezic:2005,mauroy:mezic:susuki:2020,tu2014dynamic}. Particularly successful applications encompass fluid dynamics \cite{mezic2013analysis}, molecular dynamics \cite{wu2017variational}, power systems~\cite{susuki2016applied} or coupled biological models~\cite{mangan2016inferring}, to name a few. Koopman theory, and related data-driven approximations via (E)DMD have also been extended to control systems~\cite{proctor:brunton:kutz:2018,williams:hemati:dawson:kevrekidis:rowley:2016,bevanda:sosnowski:hirche:2021,otto2021koopman}, thereby serving as a basis for data-driven controllers~\cite{bruder:fu:gillespie:remy:vasudevan:2021,budivsic2012applied}, cf.\ also for robust feedback design~\cite{strasser:schaller:worthmann:berberich:allgower:2023} or data-driven predictive control~\cite{bold:grune:schaller:worthmann:2023,peitz:otto:rowley:2020,korda:mezic:2018a}. 

Besides its straightforward implementation by regression, the popularity of EDMD is also due to its strong foundation in operator theory, numerical analysis and statistical learning theory, giving rise to various error bounds. Therein, the error typically splits up into two parts: A (probabilistic) estimation error, stemming from finitely many data points and a (deterministic) projection error, which is due to considering only finitely many observable functions in the dictionary. For ordinary differential equations, estimation error bounds for ergodic sampling were provided in~\cite{mezic2022numerical} and  for i.i.d.\ sampling, bounds including estimation and projection error using finite elements were analyzed in~\cite{zhang2021quantitative}. The estimation error for stochastic differential equations, as well as control-affine systems with either ergodic or i.i.d.\ sampling was analyzed in ~\cite{nuske2023finite,schaller2023towards}. Estimation error bounds in a general setting of dynamical systems on Polish spaces were recently provided in~\cite{philipp2024extended}. %
For dictionaries embedded in a Reproducing Kernel Hilbert Space (RKHS) and corresponding Kernel EDMD, approximation error bounds for autonomous systems were provided in~\cite{philipp2023error} and extended to control systems in~\cite{philipp2024error}. Whereas the previously mentioned bounds on the projection error (and consequently the full approximation error) were provided in $L^2$-norms, uniform, i.e., pointwise error bounds were recently proven in~\cite{kohne2024infty} using the close correspondence of regression in an RKHS norm with interpolation. The pointwise bounds were obtained using dictionaries of compactly supported Wendland kernels which are well-known to generate RKHS of (fractional) Sobolev spaces. This choice of kernel allowed for a suitable functional analytic framework as Sobolev spaces may be shown to be invariant under the Koopman operator under mild assumptions of regularity of the flow. Also recently, an EDMD-variant based on interpolation with Bernstein polynomials with corresponding pointwise error bounds was proposed in~\cite{yadav2024approximation}. %

For problems modeling interconnected systems, a common strategy in EDMD and Koopman-based approaches is to comply with and to leverage the coupling structures in the learning process. In this context, there are several works utilizing the underlying structure of the interconnection graph in EDMD.   In \cite{SchlKord22}, the authors present sparse EDMD~(sEDMD), in which  a family  of extended systems is introduced by means of the sparsity graph. %
These extended systems are self-contained and usually have smaller state space dimension than the entire system. Consequently, the Koopman operator of the whole system is decomposed into a family of operators on these smaller extended systems and approximated via EDMD. %
In \cite{jiang2024modularized} and considering an application to microgrids, each subsystem is identified  individually with a bilinear Koopman model. Then, assuming that inputs and outputs through which subsystems are connected can be determined by solving hybrid power flow equations, the subsystems are coupled.     
In \cite{Li2024parallel,mukherjee2022learning,tellez2022data},  localized EDMD (lEDMD) approaches are suggested. %
Therein, the Koopman model is learned in a distributed fashion by imposing  sparsity  on the EDMD-approximations for the entire system based on the network topology.   The central idea of these works \cite{tellez2022data,Li2024parallel,mukherjee2022learning}  is very closely related and mainly differs by the choice of observable functions.
In \cite{tellez2022data} radial basis functions are used as observables. In \cite{Li2024parallel}, the authors choose a particular kind of observable adapted to the underlying physical model such that the resulting Koopman model has a small number of states. %
The work \cite{mukherjee2022learning} proposes graph neural networks corresponding to encoders as observables.

Whereas the works mentioned in the previous paragraph pursue a learning of decoupled or local systems, there are also works that propose holistic learning architectures for the full model by exploiting structure present in subsystems or applying postprocessing in the sense of model reduction.
 In \cite{yu2023data}, a  sparsity pattern is imposed on the matrix representation of Koopman operators based on the observation that influence of a subsystems evolution is limited within a small sampling time. Then the computation of the matrix representation is reduced to  a sparse least-squares problem. In \cite{sinha2020computationally}, Cholesky decompositions are used to decrease the size of the EDMD approximant of the Koopman operator. In \cite{alla2017nonlinear}, DMD is combined with proper orthogonal decomposition to build a reduced order  nonlinear model. All these methods may  be categorized as centralized approaches, which usually, due to the curse of dimensionality, do not scale well with the  number of subsystems. %

\subsection{Contribution}
In this work, we present  the first EDMD variant for structure-exploiting modularized learning with error bounds. To achieve this, we combine two central ingredients: First, we bound the error of the local Koopman generator via existing finite-data error bounds \cite{nuske2023finite}. Second, we utilize tools from graph theory to control the propagation of these errors after coupling. More precisely, we consider (1) systems on a connected digraph without cycles, (2) weakly interconnected systems over a digraph in which the influence of neighbouring subsystems on each other is suitably bounded and (3) digraphs with cycles that allow for a condensation of strong components.

Besides the theoretical foundation via finite-data error bounds, the proposed method also has various advantages from a practical and numerical point of view. To highlight these, we provide a thorough comparison to the above mentioned lEDMD and sEDMD approaches, as well as standard EDMD. First, we will illustrate by various numerical experiments that the presented EDMD-variant is more data-efficient than existing approaches. This is due to the fact that the localization alleviates the curse of dimensionality, as observable functions are defined on the subsystems and coupled after the learning process. Second, if a model of a subsystem appears at several instances of the overall system, we can leverage this to increase the efficacy in the spirit of transfer learning. Last, if the overall system is enlarged by means of coupling another system, we only have to learn the new subsystem and in particular we do not have to retrain the whole system as would be necessary for holistic approaches.

\subsection{Organization of the article}
 In Section~\ref{sec:koop} we introduce the Koopman methodology and corresponding data-driven approximations by EDMD. Then, in Section~\ref{sec:architecture}, we propose the EDMD architecture to learn the Koopman generator for interconnected systems and a variant  learning Koopman operator. Subsequently, in Section~\ref{sec:bounds}, we deduce finite-data error bounds for the  Koopman generator. We provide various examples in Section~\ref{sec:numerics} illustrating the data-efficiency of the method. In Section~\ref{sec:transfer},  we demonstrate its advantages in transfer learning and its flexibility to topology changes. Last, we summarize in Section~\ref{sec:conclusion} the presented results and discuss some open tasks  for future works.

\subsection{Nomenclature}
 For $a,b \in \mathbb{R}_{\geq 0}$ with $a \leq b$, we write $[a:b] := [a,b] \cap \mathbb{Z}$.  Further, for $T\geq 0$, we abbreviate $\infn{x}=\sup_{t\in [0,T]}\|x(t)\|$ for  $x\in C^1([0,T], \mathbb{R}^n)$, $n\in \mathbb{N}$, with  $\| \cdot \|$ denoting the one-norm.   By $\preceq$ we  denote componentwise inequality on $\R^{m \times n}$, with $m,n \in \mathbb{N}$.  For a block matrix $\left[X_1 \ \ldots \ X_m\right] \in \R^{n \times mq }$ with $q\in \mathbb{N}$,  $X_i \in \R^{n\times q}$, we use  $ (X_i)_{i\in [1:m]} $ to abbreviate it, and use $\left((X_i)_{i\in [1:m]} \right)^{\|\cdot \|}$ to denote the  vector $\left[ \|X_1\| \ \ldots \ \|X_n\| \right]^\top \in \R^m$.   Further, we write $\langle \cdot,\cdot \rangle$ for the standard Euclidean inner product in $\R^n$, $n\in \N$.

\section{Koopman-based representation and data-driven approximation}
\label{sec:koop}
\noindent 
In this part, we briefly introduce the Koopman approach and corresponding data-driven approximations via Extended Dynamic Mode Decomposition (EDMD).

\subsection{The Koopman framework}

\noindent To provide the fundamentals of the Koopman formalism~\cite{koopman1931hamiltonian}, consider the initial value problem corresponding to a \textit{nonlinear} Ordinary Differential Equation (ODE)
\begin{align}\label{eq:ode}
    \dot x = F(x), \qquad x(0)=x^0,
\end{align}
with locally Lipschitz-continuous vector field $F: \R^n \rightarrow \R^n$ and initial value $x^0\in \R^n$, $n\in \N$.  
We denote by $x(t;x^0)$ the flow associated to the ODE~\eqref{eq:ode} at time $t \geq 0$ for %
initial state $x^0 \in \R^n$. To alleviate the exposition, we tacitly assume that the solution exists for all times $t\geq 0$. If this is not the case, one may consider the maximal interval of existence in the following considerations. 

The Koopman semigroup of operators $(\mathcal{K}^t)_{t \geq 0}$ is defined by the identity
\begin{align*}
    (\mathcal{K}^t\varphi)(x^0) = \varphi(x(t;x^0)) \qquad \forall t\geq 0,\ x^0\in \R^n,
\end{align*}
which has to hold for all observable functions $\varphi: \R^{n} \to \R$.
Due to the semigroup property of the dynamics~\eqref{eq:ode}, one may show that in various functional analytic settings, the Koopman operator forms a strongly-continuous semigroup of \textit{linear} bounded operators. In this context, we refer to \cite{philipp2024error,philipp2023error} for $L^2$-spaces and to \cite{farkas2020towards} for spaces of continuous functions. Due to strong continuity, one may define the corresponding generator by
\begin{align*}
    \qquad \mathcal{L}\varphi := \frac{\mathrm{d}}{\mathrm{d}t} (\mathcal{K}^t\varphi)\vert_{t=0} = \lim_{t\downarrow 0} \frac{\mathcal{K}^t \varphi -\varphi }{t}
\end{align*}
on a domain of definition $\operatorname{dom}(\mathcal{L})$ consisting of functions for which the above limit exists. By means of the chain rule, one may readily verify from the definition that the generator satisfies
\begin{align*}
    \calL \varphi(x^0) = \frac{\mathrm d}{\mathrm{d}t} \varphi(x(t;x^0))\big\vert_{t=0} &= \langle \nabla \varphi(x^0), F(x^0) \rangle. 
\end{align*}
for all $\varphi \in \operatorname{dom}(\calL)$ and $x^0\in \R^n$.

Recent works utilize the Koopman operator, e.g.\ for system identification, control design or stability analysis. There, the fundamental property is that, due to lifting the dynamics on an infinite-dimensional space, the Koopman operator $\mathcal{K}^t$ (or its generator $\mathcal{L}$) are linear operators. This linearity is then the basis for data-driven approximations via linear regression. The most popular technique is EDMD as briefly recapped %
in the subsequent %
subsection. %

\subsection{Data-driven approximation of the Koopman generator by EDMD}

\noindent 
In generator EDMD (gEDMD; \cite{klus:nuske:peitz:niemann:clementi:schutte:2020}), one considers a compact set $\X\subset \R^n$ and a finite-dimensional subspace of observables $\V := \spn\{\varphi^1,\ldots,\varphi^{N}\} \subset L^2(\X)$, $N\in \mathbb{N}$, with $\varphi^k: \X \to \R $ and $\varphi^k\in \operatorname{dom}(\calL)$, $k\in [1:N]$. %
Then, the compression of the Koopman generator onto the dictionary~$\V$ %
is given by $L_{\V} = P_\V \mathcal{L}\vert_{\V}$, 
 where $P_\V$ is the $L^2(\X)$-orthogonal projection onto $\V$. 
Due to finite dimensionality of the dictionary $\V$, $L_{\mathbb{V}}$ may be represented by a matrix. We denote this representation, with slight abuse of notation, by the same symbol. 

In EDMD, the matrix $L_{\V}$ is approximated by a finite number $m$ of data points $ \{x^l\}_{l\in [1:m]}$. %
To this end, we form the vector-valued observable $\Phi(x) := \begin{bmatrix}
\varphi^1(x) & \ldots & \varphi^N(x)
\end{bmatrix}^\top$ and define the data matrices
\begin{align*}
\Phi(X) := \begin{bmatrix}
\Phi(x^1) &\ldots&\Phi(x^{m})
\end{bmatrix}\in \mathbb{R}^{N\times m},\quad 
\calL\Phi(X) := \begin{bmatrix}
\mathcal{L}\Phi(x^1)&\ldots&\mathcal{L}\Phi(x^{m})
\end{bmatrix}\in \mathbb{R}^{N\times m},
\end{align*}
where $\mathcal{L}\Phi(x)$ is to be understood componentwise such that $(\mathcal{L}\Phi(x))_k = \langle \nabla \varphi^k(x),F(x)\rangle$, $k\in [1:N]$ and $x\in \X$. Then we define the EDMD approximation of the Koopman generator as a solution to the regression problem
\begin{align*}
    \operatorname{min}_{L\in \R^{N\times N}}\|\mathcal{L}\Phi(X) - L\Phi(X)\|^2_2,
\end{align*}
that is,
\begin{align*}
	L_m = (\Phi(X)\Phi(X)^\top)^{\dagger}\Phi(X) \calL\Phi(X)^\top \in \R^{N\times N}, 
\end{align*} 

The following result, taken from \cite[Theorem 12]{nuske2023finite}, yields a bound for the estimation error and provides the central tool of our analysis.
\begin{lem} \label{lem:error_generator}
Let the data points $ \{x^l\}_{l\in [0:m-1]}$ be drawn independently and identically distributed (i.i.d.) w.r.t.~the uniform distribution in $\X$. Then, for any error bound $\varepsilon >0$ and probabilistic tolerance $\delta \in (0,1)$ there is a sufficient amount of data $m_0 = \mathcal{O}(\frac{N^2}{\delta \varepsilon^2})$ such that for all $m\geq m_0$
 \begin{equation*}
 \bbP\left( \| L_{\V}- L_{m}\| \leq \varepsilon \right) \geq 1-\delta.
 \end{equation*}
\end{lem}

The above result provides a probabilistic bound on the estimation error. To bound the full approximation error, one further has to bound the difference between the compression $L_{\mathbb{V}}= P_\V \mathcal{L}\vert_{\V}$ and the Koopman generator $\calL$. Clearly, this projection error depends on the choice of dictionary functions. For EDMD applied to the Koopman operator or generator, we refer to~\cite{korda2018convergence} for a dictionary formed by orthonormal functions and to \cite{zhang2021quantitative,schaller2023towards} for $L^2$ projection error bounds for finite elements. 
For kernel-based dictionaries, we mention \cite{philipp2023error} for $L^2$-error bounds and \cite{kohne2024infty} for pointwise error bounds. Using a variant of EDMD and a dictionary consisting of Bernstein polynomials, pointwise error bounds also have been presented in \cite{yadav2024approximation}. 

To streamline the presentation in this work, we will focus on the estimation error and in particular the propagation of estimation errors of decoupled systems on the full system. Similar argumentations for the full approximation error are subject to future work. Furthermore, we will provide estimation error bounds only for our modularized EDMD variant for Koopman generators. The extension to Koopman operators presented in Subsection~\ref{sec:operator}, which only approximately inherit the control affine structure, will be considered for future work. %

\section{A novel EDMD architecture for interconnected models}
\label{sec:architecture}

\noindent 
In this part, we present a learning architecture for ODEs that may be rewritten as interconnected system. We will show that leveraging the coupling structure allows to learn local Koopman models with less data requirements. 
To provide probabilistic bounds on the corresponding estimation error in Section~\ref{sec:bounds}, we %
utilize tools from graph theory, which allow us to bound the influence of %
errors resulting from a %
subsystem on %
the overall system.%

We briefly recall some common denominations~\cite{bondy2008graph} that will be utilized in the remainder. 
A directed graph (or digraph) is an ordered pair $(\mathcal{V}, \mathcal{A})$ consisting of a set~$\mathcal{V}$ of vertices and a set~$\mathcal{A}$ of arcs associated with ordered pairs of vertices. In this work, only arcs with ordered pairs of distinct vertices are considered and each vertex is referred to as a subsystem. 
The tails (heads) of all arcs with a vertex %
as their head (tail) are called in-neighbours (out-neighbours). A directed cycle is a digraph whose vertices can be arranged in a cyclic sequence along the direction of arcs.

\subsection{Learning interconnected models via modularized generator EDMD }

\noindent In this work we consider systems given by ODEs %
of the form
\begin{align}\label{eq:globalmodel}
    \dot{x} = f(x) + G(x)x,
\end{align}
where the state $x \in \R^n$ is partitioned into $s \in \mathbb{N}$ subsystems, i.e., $x=\begin{bmatrix}
 x_1^\top & \ldots & x_s^\top   
\end{bmatrix}^\top$ with  $x_i \in \R^{n_i}$, $n_i \in \N$ for $i\in [1:s]$ and $\sum_{i=1}^s n_i = n$.  
Again, we assume that the vector field $F(x):= f(x) + G(x)x$ is locally Lipschitz continuous to ensure local existence of solutions.
We assume that the coupling is linear in the sense that the maps %
$f: \R^{n} \to \R^{n}$  and $G:\R^{n} \to \R^{n\times n}$ are given by %
\begin{align*}
    f(x) = \begin{bmatrix}
        f_1(x_1) \\
        \vdots \\
        f_s(x_s)
    \end{bmatrix},\qquad %
    G(x) = \mas{ccc}{
        G_{11}(x_1) & \ldots &  G_{1s}(x_1)\\
        \vdots &  \ddots  & \vdots\\
        G_{s1}(x_s) & \ldots  & G_{ss}(x_s) 
    },
\end{align*}
that is, the $i$-th row is only nonlinear in the $i$-th variable, $i \in [1:n]$. In the following, $\mathcal{N}_i \subseteq [1:s] \setminus \{i\}$ denotes the set of in-neighbours of the $i$-th subsystem and the matrix-valued map $G_{ij}:\R^{n_i} \to \mathbb{R}^{n_i\times n_j}$ models the impact of the respective coupling on the $i$-th subsystem. 
In particular, for $i\in [1:s]$, we may assume w.l.o.g.\ that $G_{ii}(x_i)=\mathbf{0}_{n_i\times n_i}$ in view of~$f_i(x_i)$ and further we have $G_{ij}(x_i)
\equiv \mathbf{0}_{n_i \times n_j}$ if and only if $j\notin \calN_i$. %
To motivate the proposed approach, %
let us consider subsystem $i\in [1:s]$ and the corresponding local model with state trajectory $x_i \in C^1([0,T];\R^{n_i})$ %
governed by the dynamics of %
the $i$-th row in \eqref{eq:globalmodel}, i.e., 
\begin{align}\label{eq:localmodel}
    \dot{x}_i(t) = f_i(x_i(t)) + \sum_{j \in \calN_i} G_{ij}(x_i(t)) x_j(t),  \qquad i \in [1:s].
\end{align}
Here, we already eliminated the states which are not contained in the set of in-neighbours~$\calN_i$ %
of the $i$-th subsystem as $G_{ij}(x) \equiv \mathbf{0}$ for all $j\notin \calN_i$.
To underline the dependence of ~\eqref{eq:localmodel} on the in-neighbours, for given initial value~$x_i^0 \in \mathbb{R}^{n_i}$ and continuous trajectories $x_j$, $j \in \mathcal{N}_i$, the solution of the \textit{local model}~\eqref{eq:localmodel} is denoted by $x_i(\,\cdot\,;x_i^0,x_{\calN_i})$, where we use the shorthand $x_{\calN_i} = \bigtimes_{j\in \calN_i} x_j$.

To learn a bilinear Koopman-based representation of the $i$-th, $i\in [1:s]$, %
nonlinear subsystem described by the dynamics \eqref{eq:localmodel}, we consider $N_i\in \N$ observable functions $\varphi^k_i:\ \X_i \to \R$, $k\in [1:N_i]$, defined on a compact set $\X_i$, which span the finite-dimensional subspace $\bbV_i:=\spn\{\varphi_i^1,\ldots,\varphi_i^{N_i}\}$. 
For each in-neighbour indexed by $j\in \calN_i$  with state  $x_j\in \R^{n_j}$, we denote the unit vectors in $\R^{n_j}$ by $e^{j,1},\ldots,e^{j,{n_j}}$ and set $ e^{j,0} := \mathbf{0}_{\R^{n_j}}$. 
For $r \in [1:n_j]$, we further set %
\begin{align*}
    e^r_j:=\bigtimes_{k\in \calN_i} e_k \in \R^{\sum_{k\in\calN_i}n_k} \quad \mathrm{with} \quad e_k = 
    \begin{cases}
        e^{j,r} &\mathrm{if}\ k=j\\
        e^{k,0} &\mathrm{otherwise}.
    \end{cases}
\end{align*}
Correspondingly, for $v\in \bigcup_{j\in \calN_i} \bigcup_{r=0}^{n_j}\{ e_j^r\}$, we define the \textit{local} Koopman operator and its generator for the $i$-th, $i\in [1:s]$, subsystem by\footnote{If the $i$-th subsystem is only influenced by a part of states of its in-neighbour indexed by $j\in \calN_i$, say $[x_{j,{k_j}}, \ldots, x_{j,{n_j}}]^\top$ with $1 < k_j \leq n_j $,  then $\mathcal{K}_{i,v}^t$ and $\calL^v_i$ for $v\in  \bigcup_{r\in [k_j:n_j]\bigcup \{0\}}\{ e_j^r\}$ suffices.}
\begin{align*}
    \mathcal{K}^t_{i,v} \varphi:= \varphi(x_i(t;\cdot,v)), \qquad \mathcal{L}^v_{i} \varphi := \frac{\mathrm{d}}{\mathrm{d}t} (\mathcal{K}^t_{i,v} \varphi)\big\vert_{t=0}
\end{align*}
for all $\varphi \in \mathbb{V}_i$.

The central observation laying the foundation for our proposed method is the following: The Koopman dynamics of subsystem $i\in [1:s]$ influenced by the states of the neighboring systems $j\in \calN_i$ may be predicted by means of the Koopman generators corresponding to the unit vectors and zero, that is, $\mathcal{L}^v_i$, $v\in \bigcup_{j\in \calN_i} \bigcup_{r=0}^{n_j} \{ e_j^r\}$. This is possible due to the linear coupling terms assumed in \eqref{eq:globalmodel} and is strongly related to bilinear control approaches leveraging that control-affinity is inherited by the Koopman generator, %
cf.~\cite{surana:2016,williams:hemati:dawson:kevrekidis:rowley:2016} and \cite{nuske2023finite} for error bounds, and approximately preserved for the Koopman operator, see, e.g., \cite{peitz:otto:rowley:2020,StraScha24}.
More precisely, for $i\in [1:s]$ and $\varphi \in \bbV_i$, we compute 
\begin{align*}
    (\mathcal{L}^{x_{\mathcal{N}_i}}_{i} \varphi)(x_i^0) = \frac{\mathrm d}{\mathrm{d}t} \varphi(x_i(t;x_i^0,x_{\mathcal{N}_i}))\big\vert_{t=0} &= \langle \nabla \varphi(x_i^0), f_i(x_i^0) + \sum_{j\in \calN_i}G_{ij}(x_i^0)x_j(0) \rangle \\
    &= \langle \nabla \varphi(x_i^0), f_i(x_i^0) \rangle + \sum_{j\in \calN_i} \langle \nabla \varphi(x_i^0), G_{ij}(x_i^0)x_j(0) \rangle \\
    &= (\calL^0_i\varphi)(x^0_i) + \sum_{j\in \calN_i}\sum_{r=1}^{n_j} x_{j,r}(0) \langle \nabla \varphi(x_i^0),G_{ij}(x^0_i) e^{j,r} \rangle\\
    &= (\calL^0_i\varphi)(x^0_i) + \left(\sum_{j\in \calN_i}\sum_{r=1}^{n_j} x_{j,r}(0) \left(\calL^{e_j^r}_{i} - \calL^0_i \right)\varphi \right) (x^0_i)
\end{align*}
where $x_{j,r}(0)$, $r\in [1:n_j]$, denotes the $r$-th component of $x_j(0) \in \R^{n_j}$. This means that a surrogate model of the $i$-th subsystem may be composed by means of data-driven approximations of the generators $\mathcal{L}^v_i$, $v\in \bigcup_{j\in \calN_i} \bigcup_{r=0}^{n_j}\{ e_j^r\}$.

In the remainder of this work, we assume that the dictionaries include the coordinate maps, that is, for all $i \in [1:s]$, $\varphi^k_i(x_i)=x_{i,k}$,  $k\in [1:n_i]$.
Correspondingly, we denote by $P_i$, $i\in [1:s]$, the matrix which projects the lifted state of system $i$ onto the state of system $i$, that is, we may recover the state via
  \begin{align*}
      x_i = P_i z_i, \qquad i\in [1:s].
  \end{align*}
The proposed modularized generator EDMD is summarized in the following algorithm.

\begin{algorithm}
	\caption{Modularized Generator EDMD (mgEDMD)}
	\begin{algorithmic}[1]
		\REQUIRE  Dictionaries $\V_i$, $i\in [1:s]$ including the respective coordinate maps.   %
		\FORALL { $i\in [1:s]$}
		\STATE Approximate $\calL^v_i$ by $L^v_{m_i}$ for all $v\in \bigcup_{j\in\calN_i}\bigcup_{r=0}^{n_j}\{ e_j^r\}$ using gEDMD with $m_i$ data points. 
         \STATE For $x_{\calN_i}$, define the approximation $ L^{x_{\mathcal{N}_i}}_{m_i}= L^0_{m_i}+\sum_{j\in \calN_i}\sum_{r=1}^{n_j}x_{j,r}(L^{e^r_j}_{m_i}-L^0_{m_i})$.
        \ENDFOR 
       \STATE Define bilinear Koopman model  $(\dot{z}_i)_{i\in [1:s]}= (L^{x_{\mathcal{N}_i}}_{m_i} z_i)_{i\in [1:s]}$ by setting $x_{\calN_i}= \bigtimes_{j\in \calN_i} P_j z_j$.
     \end{algorithmic}\label{Algo: mgEDMD}
\end{algorithm}

The following remark considers output couplings.

\begin{rem}
    We point out that the key property to apply our approach is the separability of the \textit{different sources} including the knowledge on the coupling structure meaning that also subsystems influenced by functions of states, e.g.\ outputs, of in-neighbours can be easily covered. To see this, consider the local models
    \begin{align}\label{eq:localmodel2}
        \dot{x}_i(t) &= f_i(x_i(t)) + \sum_{j \in \calN_i} G_{ij}(x_i(t)) y_j(t), \\  y_i(t) &=h_i(x_i(t)), 
    \end{align}
    for $i \in [1:s]$, where the maps $h_i: \R^{n_i} \rightarrow \R^{q_i}$, $ i \in [1:s]$, are supposed to be differentiable. By defining $\tilde{x}_i=[x_i^\top \ y_i^\top]^\top \in \R^{n_i+q_i}$, the local models (\ref{eq:localmodel2}) can be rewritten as 
    \begin{align}\label{eq:localmodel3}
        \dot{\tilde{x}}_i(t) = \tilde{f}_i(\tilde{x}_i(t)) + \sum_{j \in \calN_i}  \mas{cc}{\mathbf{0}_{(n_i+q_i )\times n_j } & \tilde{G}_{ij}(\tilde{x}_i(t))}\tilde{x}_j(t), \qquad i \in [1:s], 
    \end{align}
    with $\tilde{f}_i: \begin{bmatrix}
      x_i(t)^\top &  y_i(t)^\top  
    \end{bmatrix}^\top \mapsto \begin{bmatrix}
    f_i(x_i(t))^\top \    f_i(x_i(t))^\top \nabla h_i (x_i(t))     
    \end{bmatrix}^\top$, $i\in [1:s]$, and the matrix-valued maps
    $\tilde{G}_{ij}:  \begin{bmatrix}
      x_i(t)^\top & y_i(t)^\top  
    \end{bmatrix}^\top \mapsto \begin{bmatrix}
     G_{ij}(x_i(t))^\top &  G_{ij}(x_i(t))^\top \nabla h_i(x_i(t))   
    \end{bmatrix}^\top$, $i\in [1:s], j\in \calN_i$.   
\end{rem} 
While the data efficiency will clearly be showcased in the numerical experiments of Section~\ref{sec:numerics},
we briefly motivate this efficiency of the mgEDMD as defined in Algorithm~\ref{Algo: mgEDMD} approach over EDMD applied to the full system in view of data requirements. 
To this end, we consider the worst case for our approach, i.e., we suppose that each subsystem is an out-neighbour for all other subsystems, i.e., $\mathcal{N}_i = [1:s] \setminus\{i\}$ and we assume for simplicity $n_i = n_0 \in \N$ for all $i\in [1:s]$ such that $n=sn_0$.
In the proposed method,  for all $i\in [1:s]$, we have to learn $\calL^0_i$ and $\calL^{e_j^r}_{i}$, $r\in [1:n_0]$, $j\in \mathcal{N}_i$. Thus, for each $i\in [1:s]$, $1+|\mathcal{N}_i|=1+sn_0$ generators are to be approximated on state spaces of dimension~$n_0$. Conversely, if we apply standard EDMD \cite{williams2015data} to the full system as presented in Section~\ref{sec:koop}, we have to approximate one generator on the dimension of the full state $n=sn_0$.
As the data requirements scale exponentially in the state dimension (cf.\ e.g., \cite{schaller2023towards, zhang2021quantitative,kohne2024infty}) due to the curse of dimensionality we have for the proposed method data requirements of order $\mathcal{O}(s (1+sn_0) e^{n_0})$: 
This complexity comes from approximating $s$ subsystems and each we have to approximate $1+sn_0$ generators on the space dimension $n_0$, the latter leading to exponential data requirements $e^{n_0}$. In standard EDMD, however, where we have to approximate the system of state space dimension $n = sn_0$, we obtain data requirements of order $\mathcal{O}(e^{n}) = \mathcal{O}(e^{n_0 s})$. 
Hence, the data requirements of the proposed method only scale quadratically in the number of subsystems $s$, whereas the data requirements of standard EDMD scale exponentially.
Note that this does not take into effect speed-ups of the learning phase in view of parallelization of learning the generators of the subsystems.

\subsection{Modularized EDMD using the Koopman operator} %
\label{sec:operator}
\noindent In this part, we provide a brief extension of our method to the Koopman operator and refer to the extended method as  modularized EDMD (mEDMD).   To this end, in virtue of the Euler method on the continuous-time model expressed with Koopman generator above, we get, for $i \in [1:s]$ and $\varphi \in \bbV_i$,
\begin{align*}
\varphi(x_i(\Delta_t;x_i^0,x_{\calN_i})) &\approx (\varphi+ \Delta_t \calL^0_i\varphi)(x^0_i) + \left(\sum_{j\in \calN_i}\sum_{r=1}^{n_j} x_{j,r}(0) \left( \Delta_t\calL^{e^r_j}_{i}\varphi - \Delta_t \calL^0_i \varphi \right) \right) (x^0_i)  \\
      &\approx \varphi(x_i(\Delta_t;x_i^0,0)) + \left(\sum_{j\in \calN_i}\sum_{r=1}^{n_j} x_{j,r}(0) \left( \Delta_t\calL^{e^r_j}_{i}\varphi - \Delta_t \calL^0_i \varphi \right) \right) (x^0_i) \\
      &= \varphi(x_i(\Delta_t;x_i^0,0)) + \sum_{j\in \calN_i}\sum_{r=1}^{n_j} x_{j,r}(0) \left( \varphi +\Delta_t\calL^{e^r_j}_{i}\varphi - \varphi- \Delta_t \calL^0_i \varphi \right) (x^0_i) \\ 
      & \approx \varphi(x_i(\Delta_t;x_i^0,0)) +  \sum_{j\in \calN_i}\sum_{r=1}^{n_j} x_{j,r}(0) \left( \varphi(x_i(\Delta_t;x_i^0,e_j^r))- \varphi(x_i(\Delta_t;x_i^0,0)) \right) \\ 
      &= (\mathcal{K}^{\Delta_t}_{i,0}\varphi)(x^0_i) + \left(\sum_{j\in \calN_i}\sum_{r=1}^{n_j} x_{j,r}(0) \left(\mathcal{K}^{\Delta_t}_{i,e_j^r} - \mathcal{K}^{\Delta_t}_{i,0} \right)\varphi \right) (x^0_i) .
\end{align*}
Let $j_k$ be the $k$-th element of the ordered set $\calN_i$ with $s_i = |\calN_i|$,  $x_i(l)=x_i(l \Delta_t; x_i^0, x_{\calN_i})$,   
$\bar{\varphi}_i (x_i)=\begin{bmatrix}
  x_i^\top &  \varphi^{1}_i(x_i) & \ldots & \varphi^{N_i-n_i}_i(x_i)  \end{bmatrix}^\top$
and $ \mathcal{K}^{\Delta_t}_{i,j} =  \left( \mathcal{K}^{\Delta_t}_{i,e_j^r}-\mathcal{K}^{\Delta_t}_{i,0}  \right)_{r \in [1:n_j]}$. Then, the finite-dimensional approximation of the Koopman operators  $\mathcal{K}^{\Delta_t}_{i,0} $ and $\mathcal{K}^{\Delta_t}_{i,j_k} $, denoted by $(K_{i,0}^{\Delta_t})^\top$ and  $ (K_{i,j_k}^{\Delta_t})^\top$ for $k=[1:s_i]$ respectively,  are given by 
\begin{equation*}
 \begin{bmatrix}
 K_{i,0}^{\Delta_t} &  K_{i,j_1}^{\Delta_t} & \ldots &  K_{i,j_{s_i}}^{\Delta_t}    
 \end{bmatrix}= \pmb{Y}_i\pmb{X}_i^\top(\pmb{X}_i \pmb{X}_i^\top)^\dagger   
\end{equation*}
 with 
\begin{align*}
    \pmb{X}_i&=\mas{ccc}{{\bar{\varphi}}_{i}(x^{1}_i(0)) &  \ldots  & \bar{\varphi}_{i}(x^{m_i}_i(0))\\
x^{1}_{j_1}(0)\otimes \bar{\varphi}_{i}(x^{1}_i(0)) & \ldots & x^{m_i}_{j_1}(0)\otimes \bar{\varphi}_{i}(x^{m_i}_i(0)) \\
\vdots &  & \vdots   \\
x^{1}_{j_{s_i}}(0)\otimes \bar{\varphi}_{i}(x^{1}_i(0)) & \ldots & x^{m}_{j_{s_i}}(0)\otimes \bar{\varphi}_{i}(x^{m}_i(0)) 
}, \\
 \pmb{Y}_i&=\mas{ccc}{{\bar{\varphi}}_{i}(x^{1}_i(1)) &  \ldots & \bar{\varphi}_{i}(x^{m_i}_i(1))},
 \end{align*}
where $m_i$ is the number of i.i.d\ generated snapshot pairs for the $i$-th subsystem on some compact set $\X_i$. Letting $z_i(l)=\bar{\varphi}_i (x_i(l))$,  the lifted discrete-time model for each subsystem is  described by
\begin{equation} \label{eq:mEDMD_operator}
    z_i(l+1) = K_{i,0}^{\Delta_t} z_i(l) + \sum_{j \in \calN_i} K_{i,j}^{\Delta_t} (\begin{bmatrix}
     I_{n_j} & \mathbf{0}_{n_j \times (N_j-n_j)}   
    \end{bmatrix}z_j(l) \otimes z_i(l)).
\end{equation}
 The Koopman operator %
 can be also  approximated modularly with data, which allows efficient construction of Koopman model if part of the subsystem's model is known as prior. To illustrate this, let's suppose that the $j$-th subsystem  is  connected to   the identified $i$-subsystem as its new in-neighbours, then, for $i$-th subsystem,  only $\mathcal{K}_{i, j}^{\Delta_t}$ needs to be determined and can be approximated by  $(\pmb{X}_{i,j} \pmb{X}_{ij}^\top)^\dagger\pmb{X}_{i,j}\pmb{Y}_{i,j}^\top$ with  the following data matrices
\begin{align*}
    \pmb{X}_{i,j}&=\mas{ccc}{
x^{1}_{j}(0)\otimes \bar{\varphi}_{i}(x^{1}_i(0)) & \ldots & x^{m_i}_{j}(0)\otimes \bar{\varphi}_{i}(x^{m_i}_i(0)) }, \\
 \pmb{Y}_{i,j}&=\mas{ccc}{{\bar{\varphi}}_{i}(x^{1}_i(1))-d_i^{1}  &  \ldots & \bar{\varphi}_{i}(x^{m_i}_i(1))-d_i^{m_i}},
 \end{align*}
 with  $ d_i^{m}=K_{i,0}^{\Delta_t}{\bar{\varphi}}_{i}(x^{m}_i(0))-\sum_{j\in \calN_i \setminus \{ j \}} K_{i,j}^{\Delta_t} (x^{m}_{j}(0)\otimes \bar{\varphi}_{i}(x^{m}_i(0))) $.

\section{Finite-data error bounds for modulized generator EDMD}
\label{sec:bounds}

\noindent When  predicting the state trajectory over finite time horizon $t\in[0, T]$ in the Koopman framework using EDMD, an error occurs due to using only finitely many observable functions (projection error) and using only finitely many data points (estimation error). In this work, we provide an error bound on the estimation error for the proposed architecture of Algorithm~\ref{Algo: mgEDMD}. In \cite{zhang2021quantitative}, \cite{nuske2023finite},  finite-data bounds on the estimation error have been provided for a single deterministic or stochastic  ODE system. These error bounds also have been extended to control-affine systems using the control affinity of the Koopman generator, cf.~\cite{nuske2023finite,philipp2023error}. Here, we follow a similar strategy and consider the in-neighbours as input to the subsystems. However, in the full system, these virtual inputs of the subsystem are not given a priori and thus are also subject to an error. Hence, we provide an analysis using tools from graph theory to bound the influence of the corresponding discrepancy of the (virtual) inputs of the subsystems. %

 In the following, we consider a fixed subsystem $i\in [1:s]$.
 For given trajectories of the neighboring systems in the lifted space  $z_j$ and $\widehat{z}_j$, which are projected to the original state via  $x_j=P_j z_j$ and $\widehat{x}_j = P_j \widehat{z}_j$, $j\in \mathcal{N}_i$, the local unknown  bilinear Koopman model subject to the compression and its data-driven approximation are given by
 \begin{subequations}\label{eq:koopmodel}
     \begin{align}
    \dot z_i(t) &= L^{x_{\calN_i}}_{\V_i}(t)z_i(t)= \left ( L^0_{\V_i} + \sum_{j\in \calN_i}\sum^{n_j}_{r=1}x_{j,r}(t)(L^{e^r_j}_{\V_i}-L^{0}_{\V_i} )  \right ) z_i(t),  \label{eq:koopmodel_1} \\
    z_i(0) &=z^0_i \label{eq:koopmodel_2} %
    \end{align}
 \end{subequations}
    and
    \begin{subequations}\label{eq:approx_koopmodel}
            \begin{align} 
    \dot{\widehat{z}}_i(t)&=L^{x_{\calN_i}}_{m_i}(t)\widehat{z}_i(t)= \left ( L^0_{m_i} + \sum_{j\in \calN_i}\sum^{n_j}_{r=1}\widehat{x}_{j,r}(t)(L^{e^r_j}_{m_i}-L^{0}_{m_i} ) \right ) \widehat{z}_i(t), \label{eq:approx_koopmodel_1} \\
    \widehat{z}_i(0) &=z^0_i. \label{eq:approx_koopmodel_2}%
\end{align}
    \end{subequations}

 \begin{lem} \label{lem: implicit_bound}
 Let $i\in [1:s]$ and $x_j\in C([0,T];\R^{n_j})$ with $x_j(t)\in \X_j$ for all $t\in [0,T]$ and $j\in \calN_i$ be given. Further let $x_i^0\in \X_i$ be given such that the local solution defined in \eqref{eq:localmodel} satisfies $x_i(t;x^0_i,x_{\calN_i})\in \X_i$ for all $t\in [0,T]$. Set $z_i^0= \varphi_i(x_i(0))$ in \eqref{eq:koopmodel_2} and \eqref{eq:approx_koopmodel_2}. 
  Then there exist continuous monotone\footnote{A multivariable function $f: \R^n_{\geq 0} \rightarrow \R_{\geq 0}$ is monotone  if  $x\preceq y \Rightarrow f(x) \leq f(y)$.} multivariable functions  $E_i$  %
  and $E_{ij}$, such that the finite-data estimation error is bounded by %
 \begin{equation} \label{eq:bound_e}
      \infn{x_i-\widehat{x}_i} \leq   E_i \left( \| \Delta L^0_{m_i}\|, \left( (\Delta L_{m_i}^{e_j})_{j\in \calN_i} \right)^{\|\cdot\|} \right) + \sum_{j\in \calN_i} \infn{x_j-\widehat{x}_j} E_{ij} \left(\| \Delta L^0_{m_i}\|, \left( (\Delta L_{m_i}^{e_j})_{j\in \calN_i} \right)^{\|\cdot\|}\right)
 \end{equation}
 with $\Delta L^*_{m_i} = L^*_{\V_i} - L^*_{m_i}$,  $L^{e_j}_{\star} = ( L^{e^r_j}_{\star})_{r\in [1:n_j]}$ for  $*\in \{0,e_j\,|\,j\in \mathcal{N}_i\}$,  $\star \in \{ \V_i, m_i\,|\,  i \in [1:s]\}$. Furthermore,   $E_i(0,\mathbf{0})=0$ and
 \begin{equation}\label{eq: E_ij_zero}
  E_{ij}(0,\mathbf{0})=\|P_i\|\|z_i(0)\|T\exp{\left( \nu_i T  \right)} \sum^{n_j}_{r=1}( \| (L^{e^r_j}_{\V_i}-L^0_{\V_i}) \|),    
 \end{equation} 
   where $\nu_i:= 2\| L^0_{\V_i}\|+ 2\sum_{j\in\calN_i} ( \max_{x\in \X_j}\|x\|  ) \left(\| L^0_{\V_i}\| +\sum_{r=1}^{n_j}\| L^{e^r_j}_{\V_i} \| \right)$.
  \end{lem}

\begin{proof}
 As the state $x_i$ of \eqref{eq:koopmodel} is contained in a compact set by assumption, we may estimate $ \infn{x_i}\leq \max_{x\in \X_i} \|x\| =: \alpha_i$.
 Then, abbreviating $\Delta L^{e_j^r}_{m_i} = L_{\mathbb{V}}^{e^r_j} - L_{m_i}^{e^r_j}$, $r\in [1:n_j]$, $j\in \mathcal{N}_i$, we compute %
 \begin{align*}
     \|( L^{x_{\calN_i}}_{\V_i}-L^{x_{\calN_i}}_{m_i})(t)\| &\leq \| \bigg(1-\sum_{j\in \calN_i}\sum^{n_j}_{r=1} x_{j,r}(t)\bigg)\Delta L^0_{m_i} + \sum_{j\in \calN_i}\sum^{n_j}_{r=1} x_{j,r}(t) \Delta L^{e^r_j}_{m_i}\|\\
     & \qquad + \| \sum_{j\in \calN_i}\sum^{n_j}_{r=1} (x_{j,r}(t)-\widehat{x}_{j,r}(t)) ( L^{e^r_j}_{m_i}-L^{0}_{m_i} ) \| \\
     &\leq   (1+\sum_{j\in \calN_i} \infn{ x_j}) \| \Delta L^0_{m_i} \| + \sum_{j\in \calN_i} \infn{ x_j} \sum_{r=1}^{n_j}\| \Delta  L^{e^r_j}_{m_i} \|   \\
     &\qquad + \sum_{j\in \calN_i} \infn{x_j-\widehat{x}_j}\underbrace{\sum^{n_j}_{r=1}( \| \Delta L^{e^r_j}_{m_i} \| +\| \Delta L^0_{m_i}\| + \| (L^{e^r_j}_{\V_i}-L^0_{\V_i}) \|)}_{=: E^{\Delta}_{ij}\left( \| \Delta L^0_{m_i}\|, \left( (\Delta L_{m_i}^{e_j})_{j\in \calN_i} \right)^{\|\cdot\|}\right)}\\
     &\leq \underbrace{  (1+\sum_{j\in \calN_i} \alpha_j ) \| \Delta L^0_{m_i} \| + \sum_{j\in \calN_i} \alpha_j \sum_{r=1}^{n_j}\| \Delta  L^{e^r_j}_{m_i} \|      }_{=:E^{\Delta}_i\left(\| \Delta L^0_{m_i}\|, \left( (\Delta L_{m_i}^{e_j})_{j\in \calN_i} \right)^{\|\cdot\|} \right)  }+ \sum_{j\in \calN_i} \infn{x_j-\widehat{x}_j}E^{\Delta}_{ij}. 
 \end{align*}
By definition, both functions $E^{\Delta}_{ij}$ and $E^{\Delta}_i$ are continuous and  monotone, where $E^{\Delta}_i(0, \mathbf{0})=0$.  Further, we have %
  \begin{align*}
       \exp{ \left( \int^t_0\| L^{x_{\calN_i}}_{\V_i}(s)\| \romd{s} \right ) }   &\leq \exp{ \bigg ( t  (\| L^0_{\V_i}\|(1+\sum_{j\in\calN_i}\infn{x_j} ) + \sum_{j\in\calN_i}\infn{x_j} \sum_{r=1}^{n_j}\| L^{e^r_j}_{\V_i} \| \bigg)} \\
       & \leq  \exp{ \bigg ( t  \underbrace{(\| L^0_{\V_i}\|(1+\sum_{j\in\calN_i}\alpha_j ) + \sum_{j\in\calN_i}\alpha_j \sum_{r=1}^{n_j}\| L^{e^r_j}_{\V_i} \| )}_{=: \rho_i \geq 0}  \bigg)}
 \end{align*}
We have $ \infn{\widehat{x}_i}\leq \max_{x\in \X_i} \|x\| = \alpha_i$ by assumption.  To deduce a bound on the lifted estimated state $\widehat{z}_i$, we compute
 \begin{align*}
     \|\widehat{z}_i(t) \| &\leq \| z_i(0)\| \exp{\left( \int_0^{t}\| L^{x_{\calN_i}}_{m_i}(s)\| \romd{s} \right)} \\ 
       &\leq \|z_i(0)\|\exp \bigg( t\underbrace{\big((\| \Delta L^0_{m_i}\| +\| L^0_{\V_i}\| )(1\!+\!\!\sum_{j\in\calN_i}\!\alpha_j ) \!+\!\!\sum_{j\in \calN_i}\! \alpha_j\sum_{r=1}^{n_j} (\| \Delta L^{e^r_j}_{m_i} \|+\|L^{e^r_j}_{\V_i}\| )\big)}_{ =: E_{\V_i}\left(\| \Delta L^0_{m_i}\|, \left( (\Delta L^{e_j}_{m_i})_{j\in \calN_{i}} \right) ^{\|\cdot \|} \right) } \bigg). 
\end{align*}
 where the multivariable function $ E_{\V_i}$  is also continuous and  monotone. 
Gathering the bounds above, we get %
\begin{align*}
\|x_i(t)-\widehat{x}_i(t) \|&=
    \| P_i (z_i(t)-\widehat{z}_i(t))\| \leq \|P_i\| \|z_i(t)-\widehat{z}_i(t) \|   \\
    & = \|P_i\| \|  \int^t_0 \left( L^{x_{\calN_i}}_{\V_i}(s)-L^{x_{\calN_i}}_{m_i}(s)\right) \widehat{z}_i(s) + L^{x_{\calN_i}}_{\V_i}(s)(z_i(s)-\widehat{z}_i(s)) \romd{s}   \| \\
    &  \leq   \int^t_0 \|P_i\| \| (   L^{x_{\calN_i}}_{\V_i}(s)-L^{x_{\calN_i}}_{m_i}(s) ) \|\|\widehat{z}_i(s)\| + \|L^{x_{\calN_i}}_{\V_i}(s)\| \|P_i\| \|(z_i(s)-\widehat{z}_i(s))\| \romd{s}.
\end{align*}
Thus, applying the Gronwall-Bellman inequality (cf. \cite{PACHPATTE1973758}), we obtain
\begin{align*}
    \|x_i(t)-\widehat{x}_i(t) \|
    &\leq \exp{ \left( \int^t_0\| L^{x_{\calN_i}}_{\V_i}(s)\| \romd{s} \right ) } \int_0^t \|P_i\| \|( L^{x_{\calN_i}}_{\V_i}-L^{x_{\calN_i}}_{m_i})(s) \|\|\widehat{z}_i(s)\| \romd{s} \\
    & \leq \exp(\rho_i t) ( E^{\Delta}_i+\sum_{j\in \calN_i} \infn{x_j-\widehat{x}_j} E^{\Delta}_{ij} ) \|P_i\| \|z_i(0)\| t \exp(E_{\V_i} t),  
\end{align*}
with  continuous and monotone multivariable  functions $ E^{\Delta}_i$,  $E^{\Delta}_{ij}$  and  $E_{\V_i}$. 
Consequently, taking the maximum over $t\in [0,T]$, we get
\begin{align*}
     \infn{x_i-\widehat{x}_i} &\leq  \underbrace{\|P_i\| \|z_i(0)\| T }_{=:\eta_i} \exp((\rho_i+E_{\V_i}) T) \big( E^{\Delta}_i+\sum_{j\in \calN_i} \infn{x_j-\widehat{x}_j} E^{\Delta}_{ij} \big)\\
     & \leq \underbrace{  \eta_i \exp((\rho_i+E_{\V_i}) T)  E^{\Delta}_i}_{=:E_i}+ \sum_{j\in \calN_i} \infn{x_j-\widehat{x}_j} \underbrace{\eta_i\exp((\rho_i+E_{\V_i}) T) E^{\Delta}_{ij}}_{=:E_{ij}}, 
     \end{align*}
Clearly, by definition,  $E_i$ and $E_{ij}$ are continuous and  monotone with $E_i(0, \mathbf{0})=0$, and  \eqref{eq: E_ij_zero} holds with $\nu_i=\rho_i+E_{\V_i}(0,\mathbf{0})$.  
\end{proof}

When combined to an error bound on the full system by summing up the estimate of Lemma~\ref{lem: implicit_bound} over all subsystems $i\in [1:s]$, the resulting bound is of implicit nature, as the upper bound in \eqref{eq:bound_e} depends on the state error of the other systems. Thus, we will now consider three classes of systems in which we can provide explicit error bounds. The first class analysed in Subsection~\ref{subsec:acyclic} is given by systems defined on  a connected directed digraph without cycles, such that, via a topological sort,  we can guarantee  a particular structure of in-neighbours. Then, in Subsection~\ref{subsec:weaklycoupled}, we investigate the case of weakly interconnected systems over an arbitrary digraph, which loosely speaking ensures that the influence of the  in-neighbours onto the subsystem of consideration is small.  Last, in Subsection~\ref{subsec:vertex-disjoint}, we consider   a  digraph containing  circles, for which we may extract a condensation encoding the strong component of the digraph.

\subsection{Error bounds for systems on acyclic digraphs}
\label{subsec:acyclic}

\noindent 
In this subsection, we derive bounds for systems defined on acyclic digraphs. %
To begin, we first clarify some concepts and then show how to exploit the graph structure to derive bounds.     
An acyclic digraph is a digraph having no directed cycle. Since every acyclic digraph admits a topological sort (cf. \cite[ Exercise 2.1.11]{bondy2008graph}), we can derive  a linear ordering of vertices of the digraph such that, for every arc, its tail precedes its head in the ordering.
 More precisely, for the system \eqref{eq:globalmodel} defined on an acyclic digraph,   let $( v_1, v_2, \ldots, v_s )$ be its topological sort  with $v_i \in [1:s]$, then $\calN_{v_1} = \emptyset$ and  all vertices in $\calN_{v_i}$  for $i\in [2:s]$ are either contained in  $ \{v_1, \ldots, v_{i-1}\}$ or $\calN_{v_i}=\emptyset$. 

\begin{prop} \label{prop:acyclic}
 Suppose that the system \eqref{eq:globalmodel} is defined on an acyclic digraph. Let $i\in [1:s]$ and $x_j\in C([0,T];\R^{n_j})$ with $x_j(t)\in \X_j$ for all $t\in [0,T]$ and $j\in \calN_i$ be given. Further let $x_i^0\in \X_i$ be given such that the local solution defined in \eqref{eq:localmodel} satisfies $x_i(t;x^0_i,x_{\calN_i})\in \X_i$ for all $t\in [0,T]$.
 Set $z_i^0= \varphi_i(x_i(0))$ in \eqref{eq:approx_koopmodel_1} and \eqref{eq:approx_koopmodel_2} for all $i\in [1:s]$.  
 
 Then, for any probabilistic tolerance $\delta \in (0,1)$ and error bound $\varepsilon>0$,  there exists a minimal amount of data points, such that the estimation error of mgEDMD as defined in Algorithm~\ref{Algo: mgEDMD} satisfies
\begin{align*}
    \bbP\left(\infn{x-\widehat{x}}\leq \varepsilon \right) \geq \delta. 
\end{align*}   
for $x,\widehat{x}$ defined by \eqref{eq:koopmodel} and \eqref{eq:approx_koopmodel}.
\end{prop}
\begin{proof}
We  prove the statement by induction. As the $v_1$-th subsystem has no in-neighbours, i.e. $\calN_{v_1} = \emptyset$,  we get from Lemma~\ref{lem: implicit_bound} that
\begin{align*}
    \|x_{v_1} - \widehat{x}_{v_1}\|_{\infty} \leq E_{v_1} \left(\| \Delta L^0_{m_{v_1}}\|, \left( (\Delta L^{e_j}_{m_{v_1}})_{j\in \calN_{v_1}} \right)^{\|\cdot\|}\right).
\end{align*}
As $E_{v_1}(0, \mathbf{0})=0$, the upper bound may be made arbitrarily small in probability by leveraging Lemma~\ref{lem:error_generator}.

For the $v_i$-th subsystem with $i\in [2:s]$, let us assume that the statement holds for each $v_j$-th subsystem with $j \leq i-1$, i.e. $v_j \in \calN_{v_i}$. Invoking Lemma~\ref{lem: implicit_bound} on the $v_i$-th subsystem yields 
\begin{equation*}
 \infn{x_{v_i}-\widehat{x}_{v_i}} \leq   E_{v_i}  +  \sum_{j\in \calN_{v_i}} \infn{x_j-\widehat{x}_j} E_{{v_i j}},
\end{equation*}
where the arguments of the  functions $E_{v_i}$ and $E_{v_i j}$ are omitted for brevity. 

 The  probabilistic bounds  on  each $E_{v_i}$ and  $\infn{x_j-\widehat{x}_j}$ with  $j\in \calN_{v_i}$ are  arbitrarily small for sufficient amounts of data due to Lemma~\ref{lem:error_generator} and as the statement holds for the previous subsystems.  Similarly,  for any probabilistic bound $\varepsilon_{v_i j}$ larger than $E_{v_i j}(0, \mathbf{0})$  with $j\in \calN_{v_i}$, there exists amounts of data such that  $E_{v_ij}$ is  bounded by $\varepsilon_{v_i j}$ in probability in virtue of Lemma~\ref{lem:error_generator}.     Therefore, for  any $\varepsilon > 0$, $\delta \in (0,1)$,   there exists amounts of data such that $  \bbP\left(\infn{x_{v_i}-\widehat{x}_{v_i}}\leq \varepsilon \right) \geq \delta$.

\end{proof}

\subsection{Error bounds for weakly interconnected systems on arbitary digraphs} \label{subsec:weaklycoupled}
\noindent
In this subsection, we provide finite-data error bounds for weakly interconnected  systems defined on arbitrary digraphs using  the preliminary result presented in Lemma~\ref{lem: implicit_bound}. To this end,  we first define weakly interconnected systems. %

 \begin{ass}\label{Def:weaklycoupled}
The system~\eqref{eq:globalmodel} is  weakly interconnected in the lifted space $\bigtimes_{i\in [1:s]}\V_i$ over a prediction horizon $T$ and sets $ \X_i$, $i\in [1:s]$, that is, there exists  $\bar{\varepsilon} \in [0,1)$  such that for all $i\in [1:s]$
\begin{equation}\label{eq:weak_coupling}
\sum_{j\in \calN_{-i}} E_{ji}(0, \mathbf{0}) \leq \bar{\varepsilon},   \ \forall x_i(0)=\widehat{x}_i(0) \in  \X_i,
\end{equation}
where  $\calN_{-i}:=\{ l\in [1:s] : i \in \calN_l\}$, i.e. the set of out-neighbours of the $i$-th subsystem, and $E_{ji}(0,\mathbf{0})$ is defined in \eqref{eq: E_ij_zero}.  
\end{ass}
 Based on this assumption of weak interconnection, we may state our first error bound for mgEDMD.
  \begin{prop}\label{prop:arbitarygraph}
 Let Assumption~\ref{Def:weaklycoupled} hold, i.e., the system is weakly interconnected. Let $i\in [1:s]$ and $x_j\in C([0,T];\R^{n_j})$ with $x_j(t)\in \X_j$ for all $t\in [0,T]$ and $j\in \calN_i$ be given. Further let $x_i^0\in \X_i$ be given such that the local solution defined in \eqref{eq:localmodel} satisfies $x_i(t;x^0_i,x_{\calN_i})\in \X_i$ for all $t\in [0,T]$. %
 Set $z_i^0= \varphi_i(x_i(0))$ in \eqref{eq:koopmodel_2} and \eqref{eq:approx_koopmodel_2} for all $i\in [1:s]$.  
 
 Then, for any probabilistic tolerance $\delta \in (0,1)$ and error bound $\varepsilon>0$,  there exists a minimal amount of data points, such that the estimation error of mgEDMD as defined in Algorithm~\ref{Algo: mgEDMD} satisfies
\begin{align*}
    \bbP\left(\infn{x-\widehat{x}}\leq \varepsilon \right) \geq \delta
\end{align*}
for $x,\widehat{x}$ defined by \eqref{eq:koopmodel} and \eqref{eq:approx_koopmodel}.
      
  \end{prop}
\begin{proof}
   For the sake of readability,  we omit the arguments of the functions $E_i$, $E_{ij}$ in the proof. Summing up \eqref{eq:bound_e} over $[1:s]$ yields
    \begin{equation}\label{eq:proof2.4}
      \sum_{i\in [1:s]} \infn{x_i-\widehat{x}_i }  \leq \sum_{i\in[1:s]} E_i +\sum_{i\in [1:s]} \infn{x_i-\widehat{x}_i } \sum_{j\in \calN_{-i}}  E_{ji}. 
    \end{equation}
    The system is weakly interconnected by assumption and the continuous multivariable function $\sum_{j\in \calN_{-i}}E_{ij}$ is monotone. Therefore,  for any local tolerance $\bar{\varepsilon}_i \in (\bar{\varepsilon}, 1)$ with $\bar{\varepsilon}$ specified in Assumption~\ref{Def:weaklycoupled}, there is a fixed vector $\bar{\chi}_i\succ \mathbf{0}$ such that $\sum_{j\in \calN_{-i}} E_{ji}(\chi_i) \leq \bar{\varepsilon}_i$ for $\mathbf{\chi}_i \preceq \bar{\chi}_i$. A similar statement can be drawn for the continuous monotone function $E_i$ with $E_i(0, \mathbf{0})=0$.   
     Following Lemma~\ref{lem:error_generator},  the probabilistic bounds on $\| \Delta L^0_{m_i}\|$, $\| \Delta L^{e^r_j}_{m_i} \|$  can be arbitrarily small provided that  the  number of sampling data $m_i$ is large enough. Hence,  for any    $1+\frac{\delta-1}{2s}$ with $\delta \in (0,1)$, $\varepsilon >0$, $i\in [1:s]$ and $\bar{\varepsilon}_i \in (\bar{\varepsilon}, 1)$, there exists a minimal amount of data $ m_i$ for each $i$ such that
   \begin{equation}\label{eq:proof2.42} 
        \bbP{ \left( \sum_{j\in \calN_{-i}} E_{ji} \leq   \bar{\varepsilon}_i \right) } \geq 1+\frac{\delta-1}{2s},  \ \bbP{\left( E_i  \leq \varepsilon  \right)} \geq 1+\frac{\delta-1}{2s}. \ %
    \end{equation}
   By invoking the addition rule of probability, cf.\ e.g.~\cite[Lemma 22]{nuske2023finite}, and \eqref{eq:proof2.42},  we have 
    \begin{equation*} 
        \bbP{ \left(\sum_{i\in[1:s]} E_i +  \sum_{i\in [1:s]} \infn{x_i-\widehat{x}_i }  \sum_{j\in \calN_{-i}} E_{ji} \leq s\varepsilon + \sum_{i\in [1:s]} \infn{x_i-\widehat{x}_i } \bar{\varepsilon}_i \right) } \geq \delta. 
    \end{equation*}
Together with \eqref{eq:proof2.4}, this implies  %
\begin{equation*} 
    \bbP{ \left(\sum_{i\in [1:s]} (1-\bar{\varepsilon}_i ) \infn{x_i-\widehat{x}_i } \leq s\varepsilon  \right) }\geq \delta.
\end{equation*}
Since $\bar{\varepsilon}_i <1$, we conclude
\begin{equation*} 
 \bbP{ \left( \infn{ x_i-\widehat{x}_i}  \leq (1-\bar{\varepsilon}_i )^{-1} s\varepsilon \right) } \geq \delta,   \ \forall i\in [1:s],     
\end{equation*}
where  $\varepsilon > 0$ and $\bar{\varepsilon}_i \in (\bar{\varepsilon}, 1)$ are chosen arbitrarily, which finishes the proof. 
\end{proof}
    By leveraging the graph structure,   we can establish explicit error bounds under milder conditions than  weak coupling, as shown in the upcoming %
    Subsection~\ref{subsec:vertex-disjoint}.

\subsection{Error bounds for systems on cyclic digraphs}\label{subsec:vertex-disjoint}
\begin{figure}[h]
    \centering
 \tikzset{every picture/.style={line width=0.75pt}} %

\begin{tikzpicture}[x=0.75pt,y=0.75pt,yscale=-0.7,xscale=0.7]
\draw  [line width=1.5]  (90.5,124) .. controls (90.5,110.19) and (101.69,99) .. (115.5,99) .. controls (129.31,99) and (140.5,110.19) .. (140.5,124) .. controls (140.5,137.81) and (129.31,149) .. (115.5,149) .. controls (101.69,149) and (90.5,137.81) .. (90.5,124) -- cycle ;
\draw  [line width=1.5]  (90,210) .. controls (90,196.19) and (101.19,185) .. (115,185) .. controls (128.81,185) and (140,196.19) .. (140,210) .. controls (140,223.81) and (128.81,235) .. (115,235) .. controls (101.19,235) and (90,223.81) .. (90,210) -- cycle ;
\draw  [line width=1.5]  (208,210) .. controls (208,196.19) and (219.19,185) .. (233,185) .. controls (246.81,185) and (258,196.19) .. (258,210) .. controls (258,223.81) and (246.81,235) .. (233,235) .. controls (219.19,235) and (208,223.81) .. (208,210) -- cycle ;
\draw [line width=1.5]    (233.25,150.63) -- (233.03,181) ;
\draw [shift={(233,185)}, rotate = 270.42] [fill={rgb, 255:red, 0; green, 0; blue, 0 }  ][line width=0.08]  [draw opacity=0] (15.6,-3.9) -- (0,0) -- (15.6,3.9) -- cycle    ;
\draw [line width=1.5]    (115.75,150.38) -- (115.97,181) ;
\draw [shift={(116,185)}, rotate = 269.59] [fill={rgb, 255:red, 0; green, 0; blue, 0 }  ][line width=0.08]  [draw opacity=0] (15.6,-3.9) -- (0,0) -- (15.6,3.9) -- cycle    ;
\draw  [fill={rgb, 255:red, 0; green, 0; blue, 0 }  ,fill opacity=1 ] (306.83,163.75) -- (324.63,163.75) -- (324.63,159.83) -- (336.5,167.67) -- (324.63,175.5) -- (324.63,171.58) -- (306.83,171.58) -- cycle ;
\draw  [line width=1.5]  (405.67,135.67) .. controls (405.67,121.86) and (416.86,110.67) .. (430.67,110.67) .. controls (444.47,110.67) and (455.67,121.86) .. (455.67,135.67) .. controls (455.67,149.47) and (444.47,160.67) .. (430.67,160.67) .. controls (416.86,160.67) and (405.67,149.47) .. (405.67,135.67) -- cycle ;
\draw  [line width=1.5]  (348.67,210.67) .. controls (348.67,196.86) and (359.86,185.67) .. (373.67,185.67) .. controls (387.47,185.67) and (398.67,196.86) .. (398.67,210.67) .. controls (398.67,224.47) and (387.47,235.67) .. (373.67,235.67) .. controls (359.86,235.67) and (348.67,224.47) .. (348.67,210.67) -- cycle ;
\draw  [line width=1.5]  (462.67,210.67) .. controls (462.67,196.86) and (473.86,185.67) .. (487.67,185.67) .. controls (501.47,185.67) and (512.67,196.86) .. (512.67,210.67) .. controls (512.67,224.47) and (501.47,235.67) .. (487.67,235.67) .. controls (473.86,235.67) and (462.67,224.47) .. (462.67,210.67) -- cycle ;
\draw [line width=1.5]    (447.67,153.17) -- (472.31,186.94) ;
\draw [shift={(474.67,190.17)}, rotate = 233.88] [fill={rgb, 255:red, 0; green, 0; blue, 0 }  ][line width=0.08]  [draw opacity=0] (15.6,-3.9) -- (0,0) -- (15.6,3.9) -- cycle    ;
\draw [line width=1.5]    (413.67,154.42) -- (391.82,188.55) ;
\draw [shift={(389.67,191.92)}, rotate = 302.62] [fill={rgb, 255:red, 0; green, 0; blue, 0 }  ][line width=0.08]  [draw opacity=0] (15.6,-3.9) -- (0,0) -- (15.6,3.9) -- cycle    ;
\draw [line width=1.5]    (144,210) -- (204,210) ;
\draw [shift={(208,210)}, rotate = 180] [fill={rgb, 255:red, 0; green, 0; blue, 0 }  ][line width=0.08]  [draw opacity=0] (15.6,-3.9) -- (0,0) -- (15.6,3.9) -- cycle    ;
\draw [shift={(140,210)}, rotate = 0] [fill={rgb, 255:red, 0; green, 0; blue, 0 }  ][line width=0.08]  [draw opacity=0] (15.6,-3.9) -- (0,0) -- (15.6,3.9) -- cycle    ;
\draw [line width=1.5]    (398.67,210.67) -- (458.67,210.67) ;
\draw [shift={(462.67,210.67)}, rotate = 180] [fill={rgb, 255:red, 0; green, 0; blue, 0 }  ][line width=0.08]  [draw opacity=0] (15.6,-3.9) -- (0,0) -- (15.6,3.9) -- cycle    ;
\draw  [line width=1.5]  (209,124.5) .. controls (209,110.69) and (220.19,99.5) .. (234,99.5) .. controls (247.81,99.5) and (259,110.69) .. (259,124.5) .. controls (259,138.31) and (247.81,149.5) .. (234,149.5) .. controls (220.19,149.5) and (209,138.31) .. (209,124.5) -- cycle ;
\draw [line width=1.5]    (145,124.5) -- (209,124.5) ;
\draw [shift={(141,124.5)}, rotate = 0] [fill={rgb, 255:red, 0; green, 0; blue, 0 }  ][line width=0.08]  [draw opacity=0] (15.6,-3.9) -- (0,0) -- (15.6,3.9) -- cycle    ;
\draw  [dash pattern={on 4.5pt off 4.5pt}] (80,106.55) .. controls (80,99.89) and (85.39,94.5) .. (92.05,94.5) -- (155.45,94.5) .. controls (162.11,94.5) and (167.5,99.89) .. (167.5,106.55) -- (167.5,142.7) .. controls (167.5,149.36) and (162.11,154.75) .. (155.45,154.75) -- (92.05,154.75) .. controls (85.39,154.75) and (80,149.36) .. (80,142.7) -- cycle ;
\draw  [dash pattern={on 4.5pt off 4.5pt}] (197.33,106.05) .. controls (197.33,99.39) and (202.73,94) .. (209.38,94) -- (272.45,94) .. controls (279.11,94) and (284.5,99.39) .. (284.5,106.05) -- (284.5,142.2) .. controls (284.5,148.86) and (279.11,154.25) .. (272.45,154.25) -- (209.38,154.25) .. controls (202.73,154.25) and (197.33,148.86) .. (197.33,142.2) -- cycle ;
\draw  [dash pattern={on 4.5pt off 4.5pt}] (80.33,190.22) .. controls (80.33,183.56) and (85.73,178.17) .. (92.38,178.17) -- (272.28,178.17) .. controls (278.94,178.17) and (284.33,183.56) .. (284.33,190.22) -- (284.33,226.37) .. controls (284.33,233.02) and (278.94,238.42) .. (272.28,238.42) -- (92.38,238.42) .. controls (85.73,238.42) and (80.33,233.02) .. (80.33,226.37) -- cycle ;

\draw (115.72,122.7) node  [font=\large]  {$1$};
\draw (114.5,208.5) node  [font=\large]  {$2$};
\draw (232.5,209) node  [font=\large]  {$3$};
\draw (431.89,134.36) node  [font=\large]  {$v_{1}$};
\draw (374.17,210.17) node  [font=\large]  {$v_{2}$};
\draw (488.17,209.67) node  [font=\large]  {$v_{3}$};
\draw (234.22,123.2) node  [font=\large]  {$4$};
\draw (143.5,130.4) node [anchor=north west][inner sep=0.75pt]    {$v_{2}$};
\draw (260.5,129.9) node [anchor=north west][inner sep=0.75pt]    {$v_{1}$};
\draw (260.35,215.59) node [anchor=north west][inner sep=0.75pt]    {$v_{3}$};

\end{tikzpicture}
    \caption{Condensation of a cyclic diagraph.}
    \label{fig:condentstaion_bsp}
\end{figure}
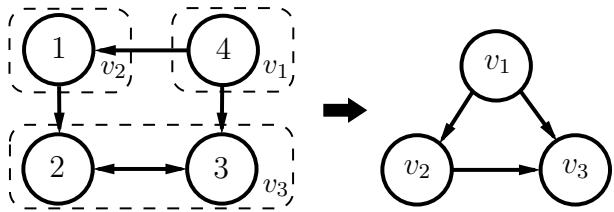

\noindent 
In this subsection, we analyse error bounds on systems defined over cyclic digraphs, i.e.\ a digraph having directed cycles. To this end, we will start by revisiting some definitions from graph theory.
A subdigraph, i.e.\ a digraph formed from subsets of the vertex set and the arc set of the original digraph,   is called a strong component  if each of its vertices is reachable from all other vertices of the total graph.
From a cyclic digraph, we can always extract a condensation, which is an acyclic digraph, whose vertices correspond to strong components, and whose two vertices are being linked by an arc if and only if there is an arc in the original digraph linking the corresponding strong components (cf. \cite{bondy2008graph}). Figure~\ref{fig:condentstaion_bsp} illustrates the condensation for a simple cyclic digraph consisting of four subsystems.

 If there is only one single vertex in the strong component, then the strong component is trivial. For a cyclic graph, at least one strong component is  non-trivial.  Suppose that the condensation extracted from the digraph of system consists of a  vertex set $\mathcal{V}$ and a set $\mathcal{A}$ of arcs.  As the condensation is acyclic (cf. \cite[Exercise 3.4.6]{bondy2008graph}), similar to Subsection~\ref{subsec:acyclic},   we can derive a topological sort  $(v_1, \ldots, v_{|\mathcal{V}|})$, $v_k \in \mathcal{V}$ from the condensation and   each vertex   $ v_k \in \mathcal{V}, k\in | \mathcal{V}|$ is assigned to  a vertex set represented by $\calI_k \subset [1:s]$.   %

\begin{ass} \label{ass:weakcoupled_2}
  Given a prediction horizon $T$ and sets $\X_i$, $i\in [1:s]$, assume that for each  non-trivial strong component $v_k$ there exists $\bar{\varepsilon}^k \in [0,1)$ such that  
\begin{equation}\label{eq:condition_new}
\sum_{j\in \calN_{-i}\bigcap \calI_k} E_{ji}(0, \mathbf{0}) \leq \bar{\varepsilon}^k,  \ \forall i\in \calI_k, \ \forall x_i(0)=\widehat{x}_i(0) \in  \X_i, 
\end{equation}
where  $\calN_{-i}$ is the set of out-neighbours of the $i$-th subsystem. 
\end{ass}
  
In contrast to the condition \eqref{eq:weak_coupling} in the definition of weakly interconnected systems, the variables $j$ and $i$ in \eqref{eq:condition_new} for each non-trivial strong component $v_k$ are specified within the subsets of $\calN_{-i}$ and $ [1:s]$ respectively. 
Therefore, \eqref{eq:condition_new} for each $v_k$ can be induced from \eqref{eq:weak_coupling}, rendering Assumption~\ref{ass:weakcoupled_2} weaker than Assumption~\ref{Def:weaklycoupled}.

\begin{prop} \label{prop:cyclic}
 Let Assumption~\ref{ass:weakcoupled_2} hold. Let $i\in [1:s]$ and $x_j\in C([0,T];\R^{n_j})$ with $x_j(t)\in \X_j$ for all $t\in [0,T]$ and $j\in \calN_i$ be given. Further let $x_i^0\in \X_i$ be given such that the local solution defined in \eqref{eq:localmodel} satisfies $x_i(t;x^0_i,x_{\calN_i})\in \X_i$ for all $t\in [0,T]$. 

 Then, for any probabilistic tolerance $\delta \in (0,1)$ and error bound $\varepsilon>0$,  there exists a minimal amount of  data points, such that the estimation error of mgEDMD as defined in Algorithm~\ref{Algo: mgEDMD} satisfies
\begin{align*}
    \bbP\left(\infn{x-\widehat{x}}\leq \varepsilon \right) \geq \delta. 
\end{align*}
for $x,\widehat{x}$ defined by \eqref{eq:koopmodel} and \eqref{eq:approx_koopmodel}.
\end{prop}
\begin{proof} We omit the arguments of functions for the sake of  brevity and proceed by induction. 

If $|\mathcal{V}|=1$, then the whole digraph is a strong component, i.e  $\calI_1=[1:s]$. Hence,  the condition \eqref{eq:condition_new}  is reduced to \eqref{eq:weak_coupling} for weakly interconnected systems.   By invoking Proposition~\ref{prop:arbitarygraph}, the claim is proven for this case. 

If $|\mathcal{V}|>1$, we first consider the strong component $v_k$ with $k>1$ and  assume that the probabilistic error bounds on  all the subsystems belonging to  the strong components $v_{j-1}$ with $j\in [2:k]$ can be arbitrarily small with sufficient amount of data. 

 If the strong component corresponding to $v_k$ is trivial, by following the same line of reasoning from the proof of Proposition~\ref{prop:acyclic}, we can show again that its error bound is arbitrarily small with sufficient amount of data. Otherwise,  summing up \eqref{eq:bound_e} over $i\in \calI_k$ yields
    \begin{equation}\label{eq:cyclic_proof_1}
      \sum_{i\in \calI_k} \infn{x_i-\widehat{x}_i }  \leq \sum_{i\in \calI_k} E_i +\sum_{i\in \calI_k} \infn{x_i-\widehat{x}_i } \sum_{j\in \calN_{-i}\bigcap \calI_k}  E_{ji}+ \sum_{i \in \calI_k }\sum_{j \in \calN_{i}\setminus \calI_k} \infn{x_j-\widehat{x}_j } E_{ij}.  
    \end{equation}
  By  the assumptions, the properties of $E_i$ and $E_{ij}$ and Lemma~\ref{lem:error_generator}, we can show that, for any $\varepsilon^k >0$, $\varepsilon^k_{i}\in (\bar{\varepsilon}^k, 1)$ with $\bar \varepsilon^k$ as in Assumption~\ref{ass:weakcoupled_2}, $\varepsilon_j>0$,  $\varepsilon^k_{ij}>E_{ij}(0, \mathbf{0})$,  there exist amounts of data such that $\sum_{i\in \calI_k}E_i$, $\sum_{j\in \calN_{-i}\bigcap \calI_k}  E_{ji}$, $\infn{x_j-\widehat{x}_j }$ and $ E_{ij}$ are probabilistically bounded by $\varepsilon^k$, $\varepsilon^k_{i}$, $\varepsilon_j$ and $\varepsilon^k_{ij}$ respectively, provided a sufficient amount of data. Thus,  from \eqref{eq:cyclic_proof_1}, we obtain
     \begin{equation*} 
 \bbP{ \left( \sum_{i\in \calI_k} \infn{x_i-\widehat{x}_i } (1-\varepsilon^k_i)  \leq \big( \varepsilon^k + \sum_{i \in \calI_k }\sum_{j \in \calN_{i}\setminus \calI_k} \varepsilon_j \varepsilon^k_{ij} \big) \right) } \geq \delta,     
      \end{equation*}
  with probabilistic tolerance $\delta \in (0,1)$. As $\varepsilon^k>0, \varepsilon_j>0, 1-\varepsilon^k_i >0$ and $\varepsilon^k_{ij}$ are chosen arbitrarily, the error bounds for $v_k$ can be arbitrarily small with enough data.
  
 Consider $k=1$ and the corresponding strong component $v_1$. Due to the topological sort, $\{i\in \calI_k: \calN_i\setminus \calI_k \neq \emptyset \}= \emptyset$. By invoking Proposition~\ref{prop:arbitarygraph} on $v_1$, the error bound for the strong component $v_1$ can be arbitrarily small with sufficient amount of data.  
\end{proof}

If no vertices in the digraph are shared by cycles, then the error bounds can be obtained under even a softer condition than Assumption~\ref{ass:weakcoupled_2} in the following sense. 
\begin{prop} \label{theo:case_2} 
Let the system \eqref{eq:globalmodel} be defined on a cyclic digraph, where no vertex is shared by at least two cycles.   Let $i\in [1:s]$ and $x_j\in C([0,T];\R^{n_j})$ with $x_j(t)\in \X_j$ for all $t\in [0,T]$ and $j\in \calN_i$ be given. Further let $x_i^0\in \X_i$ be given such that the local solution defined in \eqref{eq:localmodel} satisfies $x_i(t;x^0_i,x_{\calN_i})\in \X_i$ for all $t\in [0,T]$. 
Set $z_i^0= \varphi_i(x_i(0))$ in \eqref{eq:koopmodel_2} and \eqref{eq:approx_koopmodel_2} for all $i\in [1:s]$.
Assume that for each  non-trivial strong component $\calI_k$, there exists $\bar{\varepsilon}_k \in [0,1)$ such that  
\begin{equation} \label{eq:smallgain}
\prod_{i\in \calI_k } \left( \prod_{j\in \calN_i \cap \calI_k} E_{ij}(0, \mathbf{0}) \right) \leq \bar{\varepsilon}_k,  \forall x_i(0)=\widehat{x}_i(0) \in  \X_i. 
\end{equation} 
 Then, for any probabilistic tolerance $\delta \in (0,1)$ and  error bound $\varepsilon>0$,  there exists a sufficient amount of data such that the estimation error of mgEDMD as defined in Algorithm~\ref{Algo: mgEDMD} satisfies
\begin{align*}
   \bbP\left(\infn{x-\widehat{x}}\leq \varepsilon \right) \geq \delta. 
\end{align*}   
\end{prop}

\begin{proof} The arguments of functions are omitted for readability. The proof is conducted  by similar reasoning as in the proof of Proposition~\ref{prop:cyclic} and by exploiting the particular graph structure of non-trivial strong components.  Hence, it suffices to derive error bounds for  one non-trivial strong components $v_k$ and $|\mathcal{V}|>1$. 

As any non-trivial strong component in the graph is a circle by Lemma~\ref{Appendix:2}, one can rearrange the vertices from $\calI_k$ in a cyclic sequence along the direction of arcs. Specifically, let $(\sigma_0, \sigma_1, \ldots, \sigma_{|\calI_k|})$ be the rearranged sequence of vertices of $\calI_k$, where $ \sigma_0=\sigma_{|\calI_k|}$ and $\sigma_{i-1}$ is the in-neighbour of $\sigma_{i}$, $i\in[1:|\calI_k|]$. 
The error bound derived in Lemma~\ref{lem: implicit_bound} for  each subsystem with index   $\sigma_i \in \calI_k$  reads as   
\begin{equation*} 
\infn{x_{\sigma_i}-\widehat{x}_{\sigma_i}} \leq  \underbrace{ E_{\sigma_i}  +  \sum_ {j\in \calN_{\sigma_i} \setminus \{ \sigma_{i-1}\}} \infn{x_{j}-\widehat{x}_{j}} E_{\sigma_i j}}_{=:b_i}+ \infn{x_{\sigma_{i-1}}-\widehat{x}_{\sigma_{i-1}}} E_{\sigma_i {\sigma_{i-1}}} 
\end{equation*}
 for  $i\in [1:|\calI_k|]$, where $b_i$ is a function of the estimation errors  $\left((\Delta L^{e_j}_{m_{\sigma_i}} )_{j\in \calN_{\sigma_i} \setminus \{ \sigma_{i-1}\} }\right)^{\|\cdot\|}$ and $\|\Delta L^0_{m_{\sigma_i}}\|$.
 Denoting $e_i := \infn{x_{\sigma_i}-\widehat{x}_{\sigma_i}}$ for $i\in [1:|\calI_k|] $  and collecting the above inequalities yields  $A e\leq b$, where each entry $A_{ij}$, $i,j\in [1:|\calI_k|]$  is a function of the estimation errors, i.e.,
\begin{equation*}
    A_{ij}=
    \begin{cases}
        1, & \text{if } i=j\\
        -E_{\sigma_i \sigma_{i-1}}, & \text{if }  2 \leq i\leq |\calI_k|, j=i-1 \\
        -E_{\sigma_1 \sigma_0}, & \text{if } i=1,j=|\calI_k|\\
        0, & \text{otherwise.}
    \end{cases}
\end{equation*}
Following \eqref{eq:smallgain} and the monotony of $E_{ij}$ together with Lemma~\ref{lem:error_generator}, for any $\varepsilon_k \in (\bar{\varepsilon}_k, 1)$, $\delta \in (0,1)$,  there exists a sufficient amount of data $\bar{m}^k_{\sigma_i}\in \N$ for each subsystem belonging to  $\calI_k$ such that, for all $m^k_{\sigma_i}\geq \bar{m}^k_{\sigma_i}$, 
\begin{equation*}
    \bbP \left( \prod_{i\in \calI_k } \bigg( \prod_{j\in \calN_i \cap \calI_k} E_{ij} \bigg) \leq  \varepsilon_k \right) \geq  \dfrac{1+\delta}{2}.  
\end{equation*}
This together with non-negativity of $E_{ij}$ reveal that $A \in \mathcal{M}_{|\calI_k|}$  with probability at least $\frac{1+\delta}{2}$, where $\mathcal{M}_n$, $n\in \N$ is defined in Lemma~\ref{Appendix:1}.  Therefore,  for all $m^k_{\sigma_i} \geq \bar{m}^k_{\sigma_i}$, $i\in [1: |\calI_k|]$,   
\begin{equation} \label{proof:case_2_1}
   \bbP \left( e \preceq A^{-1} b \right) \geq \dfrac{1+\delta}{2}.
\end{equation}
Similar to Proposition~\ref{prop:cyclic}, let us consider the case $k>1$ and assume %
that the error bounds for all the subsystems with indexes in  $\calI_{j-1}$, $j\in [2:k]$,  can be arbitrarily small with enough data. 
This assumption together with properties of functions $E_i$ and $E_{ij}$ as well as Lemma~\ref{lem:error_generator} show that, for any  $\bar{b}_i >0$ with $i \in [1:|\calI_k|]$, there exist enough amounts of data $m_j$ such that $b_i \leq \bar{b}_i$ in probability. With this result in mind, we can exploit  \eqref{Appendix:monotone} from Lemma~\ref{Appendix:1} to show that, for any $\delta \in (0,1)$,  $\bar{b} = [\bar{b}_1 \ \ldots \ \bar{b}_{|\calI_k|}]$ with $\bar{b}_i >0$, and  $\Lambda \in \R_{>0}^{|\calI_k| \times |\calI_k|}$ with large enough entries, there exist fixed amounts of data  $\widehat{m}^k_{\sigma_i} \geq \bar{m}^k_{\sigma_i}$ and $\bar{m}_j$ such that 
 \begin{equation} \label{proof:case_2_2}
   \bbP \left(  A^{-1} b \preceq \Lambda \bar{b}  \right) \geq \dfrac{1+\delta}{2}, \ \forall m^k_{\sigma_i} \geq \widehat{m}^k_{\sigma_i}, \ m_j \geq \bar{m}_j.
\end{equation}
 By leveraging the addition rule of probability  together with \eqref{proof:case_2_1} and\eqref{proof:case_2_1}, we get
 \begin{equation} \label{proof:case_2_3}
     \bbP\left( e \preceq  \Lambda \bar{b}  \right) \geq \bbP \left(  
    e \preceq A^{-1} b \ \text{and} \   A^{-1} b \preceq \Lambda \bar{b} \right) \geq \delta. 
 \end{equation}
As $\bar{b}_i>0$ and $\Lambda$ can be chosen arbitrarily, we show that error bounds for the strong component $v_k$ can be arbitrarily small with enough data.

In the case of $k=1$, the strong component $v_1$ has no in-neighbours. Following the similar arguments for the case $k>1$, we can again show that all error bounds for subsystems from $\calI_1$ are arbitrarily small with enough data, which verifies the assumption   for  $k>1$.

\end{proof}

When examining a non-trivial strong component $v_k$, the condition \eqref{eq:condition_new} implies that $E_{ji}(0, \mathbf{0}) < 1$ for $i \in \calI_k$ and $j \in \calN_{-i}\bigcap \calI_k$, which is equivalent to $E_{ij}(0, \mathbf{0}) < 1$ for $i \in \calI_k$ and $j \in \calN_{i}\bigcap \calI_k$. This immediately leads to \eqref{eq:smallgain} and shows that the assumption on $E_{ij}$ in Proposition~\ref{theo:case_2} is milder than Assumption~\ref{ass:weakcoupled_2}.

\section{Numerical simulations for data-efficiency} %
\label{sec:numerics}
\noindent To show the data-efficiency  of the proposed modularized EDMD approach, we compare the prediction accuracy from the proposed approach with that from other approaches. For the comparison, the data set for learning, the type of observable functions, the overall number of states of Koopman models and the sets of initial conditions  utilized  for evaluation are the same for all the approaches. %

\subsection{Overview of existing approaches} \label{Overview}

\noindent In the following, we compare the proposed technique with  Extended Dynamic Mode Decomposition (EDMD) from \cite{williams2015data}, sparse EDMD from \cite{SchlKord22}, as well as localized EDMD from \cite{tellez2022data} and  briefly explain the fundamentals of these methods.

\noindent \textbf{Extend Dynamic Mode Decompostion (EDMD).}
Applying EDMD from \cite{williams2015data} to the entire system \eqref{eq:globalmodel}  results in  a linear predictor  for the overall system. 
 Choosing the vector-valued observable 
 $ \bar{\varphi}(x) = \begin{bmatrix}
 \varphi^{1}(x) &  \ldots & \varphi^{N}(x) 
 \end{bmatrix}^\top$ with $N \in \N$ and $\varphi^i \in \V$ defined on $\X$, 
 and letting $z(l)=\bar{\varphi}(x(l))$, we  express the predictor by
\begin{equation*}
    z(l+1)=K_{0}^{\Delta_t} z(l), 
\end{equation*}
where the matrix approximation of Koopman operator $K_{0}^{\Delta_t}$ is determined by 
\begin{align*}
    \pmb{X}&=\mas{ccc}{
\bar{\varphi}(x^{1}(0)) & \ldots & \bar{\varphi}(x^{\bar{m}}(0)) }, \\
 \pmb{Y}&=\mas{ccc}{{\bar{\varphi}}(x^{1}(1))  &  \ldots & \bar{\varphi}(x^{\bar{m}}(1))},
 \end{align*}
with $\bar{m}$ as the number of snapshots.

\noindent \textbf{Sparse EDMD (sEDMD).}
The  sparse EDMD approach proposed in \cite{SchlKord22} involves gathering all subsystems that directly or indirectly affect a specific subsystem (such as its in-neighbors and the in-neighbors of its in-neighbors) to create a self-contained extended system, on which EDMD is applied to  construct a linear predictor for each original subsystem.  
More specifically, for  the coupled Duffing system in the following Subsection \ref{CDS}, the in-neighbours of each subsystem $x_i \in \R^{2}$, $i \in [1:3]$ are  $\calN_1=\emptyset$, $\calN_2=\calN_3=\{1\}$, which induce a couple of  extended systems, whose states are $\bar{x}_1=x_1$, $\bar{x}_2=[x_1^\top \ x_2^\top]^\top$ and $\bar{x}_2=[x_1^\top \ x_3^\top]^\top$ respectively. For each extended system indexed by $i\in [1:3]$, the observables are chosen to be $\bar{\varphi}_i(\bar{x}_i)= [ \varphi_i^1(\bar{x}_i) \ \ldots \ \varphi_i^{N_i}(\bar{x}_i) ]^\top$ with $N_i \in \N$.
 Applying EDMD on each extended  system generates a predictor for each original subsystem:  
\begin{equation*}
    z_i(l+1)=K_{i,0}^{\Delta_t} z_i(l), i\in [1:s], 
\end{equation*}
where $z_i(l)=\bar{\varphi}_i(\bar{x}_i(l))$ and $K_{i,0}^{\Delta_t}$ is determined by regression with data matrices
\begin{align*}
    \pmb{X}_i&=\mas{ccc}{
\bar{\varphi}_i(\bar{x}_i^{1}(0)) & \ldots & \bar{\varphi}_i(\bar{x}_i^{m_i}(0)) }, \\
 \pmb{Y}_i&=\mas{ccc}{{\bar{\varphi}_i}(\bar{x}^{1}_i(1))  &  \ldots & \bar{\varphi}_i(\bar{x}^{m_i}_i(1))},
 \end{align*}
 with $m_i$ as the number of snapshots for the extended system associated with the  state $\bar{x}_i$. 
 
In the sparse EDMD approach, a sparse interconnection graph (or graph, from which a family of small ''subsystems"\footnote{The quotation marks are used to emphasize that in \cite{SchlKord22}, the subsystems are actually some self-contained extended systems, contrary to the subsystems used in our approach} can be drawn) is indispensable for good scalability. To see this, if all the subsystems are connected in a chain, then the subsystem in the end of the chain needs to be extended to cover the entire system, and hence the curse of dimensionality in the full dimension arises.  %

\noindent \textbf{Localized EDMD (lEDMD).}  In \cite{tellez2022data},  each individual subsystem  is locally identified by utilizing linear Koopman representation along with the prior knowledge of interconnection, i.e.\ the neighborhoods $ \calN_i$, $i\in [1:s]$. After choosing  observables functions to be 
$\bar{\varphi}_i(x_i)=[ \varphi_i^{1}(x_i)\ \ldots \ \varphi_i^{N_i}]^\top$ with $\varphi_i^k $  
for each subsystem and letting $z_i(l)=\bar{\varphi}_i(x_i(l))$, the predictor for each subsystem is given in the form,
\begin{equation} \label{eq:lDMD}
       z_i(l+1)=A_{i,0}z_i(l) + \sum_{j\in \calN_i} B_{i,j} z_j(l), i\in [1:s], 
\end{equation}
where $\begin{bmatrix}
 A_{i,0} & B_{i,j_1}& \ldots & B_{i,j_{s_i}}   
\end{bmatrix}$ are submatrices of the matrix approximation of Koopman operator with respect to the augmented observables $ \begin{bmatrix}
\bar{\varphi}_i(x_i)^\top & \bar{\varphi}_{j_1}(x_{j_1})^\top & \ldots & \bar{\varphi}_{j_{s_i}}(x_{j_{s_i}})^\top 
\end{bmatrix}^\top $. The data matrices used to approximate the predictor are as follows:  
\begin{align*}
    \pmb{X}_i&=\mas{ccc}{
\bar{\varphi}_i(x_i^{1}(0)) & \ldots & \bar{\varphi}_i(x_i^{m_i}(0)) \\
\bar{\varphi}_{j_1}(x_{j_1}^{1}(0)) & \ldots & \bar{\varphi}_{j_1}(x_{j_1}^{m_i}(0))\\
 \vdots & & \vdots \\
\bar{\varphi}_{j_{s_i}}(x_{j_{s_i}}^{1}(0)) & \ldots & \bar{\varphi}_{j_{s_i}}(x_{j_{s_i}}^{m_i}(0)) }, \\
 \pmb{Y}_i&=\mas{ccc}{{\bar{\varphi}_i}(x^{1}_i(1))  &  \ldots & \bar{\varphi}_i(x^{m_i}_i(1))}.
 \end{align*}

The main idea of lEDMD is very similar to our proposed approach. However, in our suggested learning architecture, we construct bilinear Koopman model for each subsystem instead of  linear model as in lEDMD. %

\subsection{Evaluation for systems with state and output couplings }
\noindent We provide two examples illustrating the efficacy of our approach. The first example will be given by Duffing oscillators, which are interconnected through states. The second example is given by Van-der-Pol oscillators coupled via outputs.

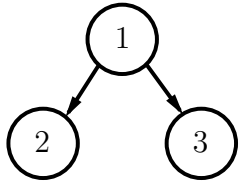
\begin{figure}[h!]
\centering

\tikzset{every picture/.style={line width=0.75pt}} %

\begin{tikzpicture}[x=0.75pt,y=0.75pt,yscale=-0.7,xscale=0.7]
\draw  [line width=1.5]  (151,135) .. controls (151,121.19) and (162.19,110) .. (176,110) .. controls (189.81,110) and (201,121.19) .. (201,135) .. controls (201,148.81) and (189.81,160) .. (176,160) .. controls (162.19,160) and (151,148.81) .. (151,135) -- cycle ;
\draw  [line width=1.5]  (94,210) .. controls (94,196.19) and (105.19,185) .. (119,185) .. controls (132.81,185) and (144,196.19) .. (144,210) .. controls (144,223.81) and (132.81,235) .. (119,235) .. controls (105.19,235) and (94,223.81) .. (94,210) -- cycle ;
\draw  [line width=1.5]  (208,210) .. controls (208,196.19) and (219.19,185) .. (233,185) .. controls (246.81,185) and (258,196.19) .. (258,210) .. controls (258,223.81) and (246.81,235) .. (233,235) .. controls (219.19,235) and (208,223.81) .. (208,210) -- cycle ;
\draw [line width=1.5]    (193,152.5) -- (217.64,186.27) ;
\draw [shift={(220,189.5)}, rotate = 233.88] [fill={rgb, 255:red, 0; green, 0; blue, 0 }  ][line width=0.08]  [draw opacity=0] (15.6,-3.9) -- (0,0) -- (15.6,3.9) -- cycle    ;
\draw [line width=1.5]    (159,153.75) -- (137.16,187.88) ;
\draw [shift={(135,191.25)}, rotate = 302.62] [fill={rgb, 255:red, 0; green, 0; blue, 0 }  ][line width=0.08]  [draw opacity=0] (15.6,-3.9) -- (0,0) -- (15.6,3.9) -- cycle    ;

\draw (176.22,133.7) node  [font=\large]  {$1$};
\draw (118.5,208.5) node  [font=\large]  {$2$};
\draw (232.5,209) node  [font=\large]  {$3$};
\end{tikzpicture}
    \caption{Interconnected Duffing oscillators }
    \label{fig:sparse_duffing}
\end{figure}

\noindent \textbf{System with state coupling: Duffing oscillators.} \label{CDS} The coupled unforced Duffing system adapted from \cite{SchlKord22} is   given by 
\begin{align*}
    \dot{x}_{1,1} &= 0.5 x_{1,2}, \ \dot{x}_{1,2} = -0.5  x_{1,2}- x_{1,1}^3,\\
    \dot{x}_{i,1} &= \alpha_i x_{i,2}, \ \dot{x}_{i,2} = -0.5 x_{i,2}-\beta_i x_{i,1}^3+\gamma_i x_{1,1}, \qquad i \in \{2,3\}, 
\end{align*}
where  $\alpha_2 = \alpha_3 = 0.5$, $\beta_1=\beta_3 = -1$ and  $ \gamma_2=0.5 \gamma_3 = 0.25$. %
The corresponding interconnection topology is depicted in Figure~\ref{fig:sparse_duffing}. Since the underlying graph of the whole system is acyclic, Proposition~\ref{prop:acyclic} specifying the prediction error of mgEDMD applies in this example. %

 We compare the  prediction errors of predictors constructed by different approaches based on $\bar{m}=1500$ and $\bar{m}=5000$ pairs of snapshots. The  initial conditions are generated randomly with uniform distributions on the box $\X=[-1.5, 1.5]^6$ for each data set. The data are collected with the sampling period  $\Delta_t=0.01$s. For sEDMD and lEDMD and  the proposed approaches,   $m_i=\bar{m}$ pairs of snapshots are extracted from samples generated by the whole system to construct data matrices for each subsystem or extended system.    The number of states of the predictor for the entire system is $N=456$.   The observables involve the thin plate RBF defined by 
    \begin{equation} \label{eq:RBF}
     \varphi^{k}(x)=\| x-c_k\|^2 \log \| x-c_k\|, \qquad k\in [1:N],   
    \end{equation} 
    with some randomly chosen center $c_k$. 
    
The configurations for all implemented approaches are summarized as follows:
\begin{itemize}

    \item \emph{EDMD:}  The observable  $\bar{\varphi}(x)$ is chosen to be $\bar{\varphi}(x) = [x^\top \  \varphi^1(x) \ \ldots \ \varphi^{N-6}(x)]^\top$, where $x=[x_1^\top \ x_2^\top \ x_3^\top]^\top$ and $\varphi^k$ is a thin plate RBF with a randomly selected center $c_k$ over  $\X = [-1.5, 1.5]^6$ for each $k$. 
    \item \emph{sEDMD}: For each induced extended system ($\bar{x}_1=x_1$, $\bar{x}_2 = [x_1^\top \ x_2^\top]^\top$ and $\bar{x}_3 = [x_1^\top \ x_3^\top]^\top$), the observable is  chosen to be  $\bar{\varphi}_i (\bar{x}_i) =  [\bar{x}_i^\top \  \varphi_i^1(\bar{x}_i) \ \ldots \ \varphi_i^{N_i-\bar{n}_i}(\bar{x}_i)]^\top$ with  $i \in [1:3]$, where $N_i=N/3$, $\bar{n}_1 = 2$,  $\bar{n}_2 = \bar{n}_3 = 4$ and  $\varphi_i^k$ is a thin plate RBF given in \eqref{eq:RBF}  with randomly chosen centers over $[-1.5, 1.5]^2$ for $\bar{x}_1$ and  $[-1.5, 1.5]^4$ for both $\bar{x}_2$ and $\bar{x}_3$. 
    \item \emph{lEDMD}:  The observable is  chosen to be  $\bar{\varphi}_i (x_i) =  [x_i^\top \  \varphi_i^1(x_i) \ \ldots \ \varphi_i^{N_i-2}(x_i)]^\top$ for each subsystem indexed by $i \in [1:3]$, where $\varphi^k_i$ is a thin plate RBF~\eqref{eq:RBF} with $c_k$ chosen randomly  in $\X_i= [-1.5, 1.5]^2$.    
     \item \emph{mgEDMD, mEDMD:} For each subsystem,  the observable $\bar{\varphi}_i$ is chosen to be the same as that for lEDMD. 
\end{itemize}

Note that in the sEDMD approach,  the extended subsystems $\bar{x}_2 \in \R^4 $ and $\bar{x}_3 \in \R^4$ have higher state space dimension than $\bar{x}_1$. As indicated by the results from \cite{zhang2021quantitative},  higher numbers of observables for  $\bar{x}_2 $ and $\bar{x}_3$ are actually necessary for a better prediction quality for  higher dimensional states. However, this will increase the overall state dimension $N$ of the predictor for the whole system after assembly.   Therefore, for the fairness of comparison,  we do not increase the number of observables for $\bar{x}_2$ and $\bar{x}_3$.

\begin{figure}[h]
\centering
\vspace{-10pt}
\begin{subfigure}{0.4\textwidth}
\hspace{75pt}
  \includegraphics[width=0.3\textwidth, clip=false, trim = 60mm 100mm 55mm 62mm]{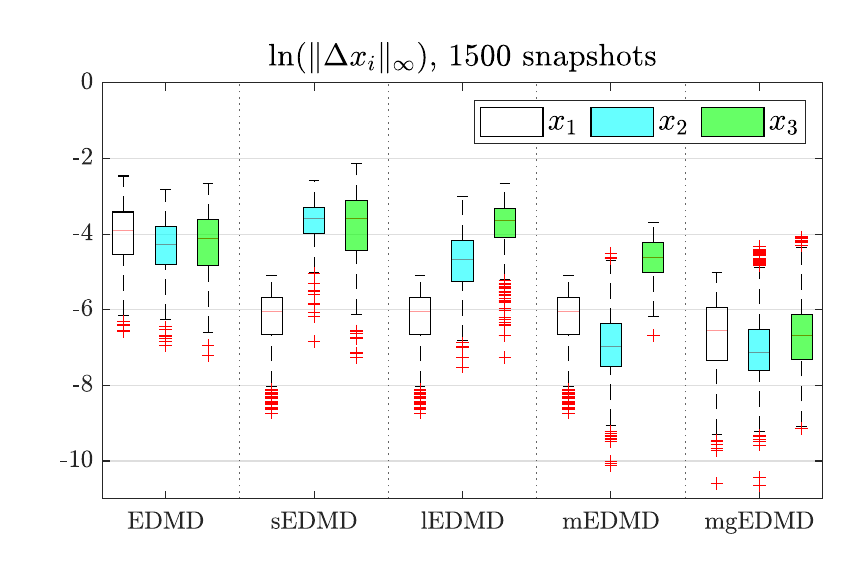}
\end{subfigure}
\hfill
\begin{subfigure}{0.4\textwidth}
\hspace{45pt}
\includegraphics[width=0.3\textwidth, clip=false, trim = 60mm 100mm 55mm 62mm]{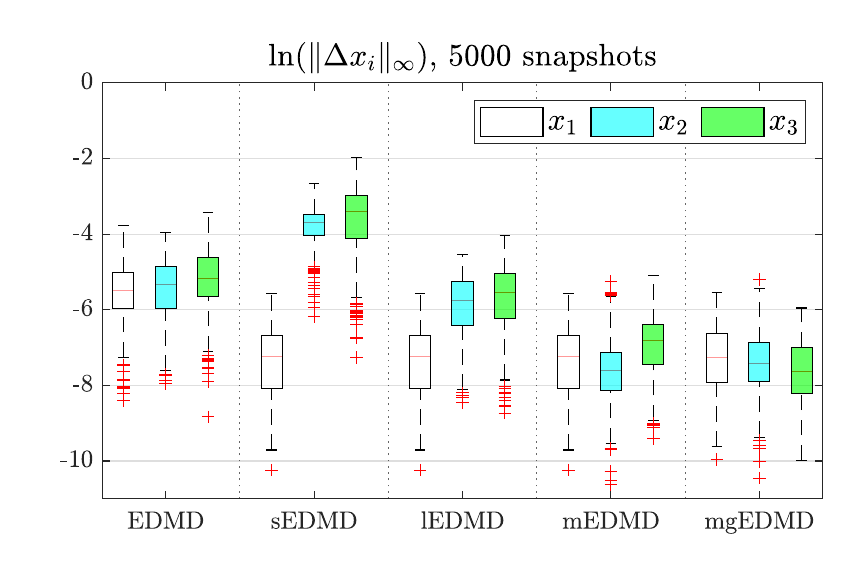}
\end{subfigure}
\vspace{155pt}
\caption{Prediction errors for Duffing oscillators with different approaches }
\label{fig:duffing_1}
\end{figure}

For the evaluation, we implement $500$ simulations with randomly chosen initial conditions in $[-0.5, 0.5]^6$. In each simulation, the trajectory is predicted over the time interval $[0, 0.5s]$ and the natural logarithm of the maximal one-norm of prediction error, denoted by $\ln(\|\Delta x_i\|_\infty)$, is computed for each predictor built by different approaches and two different data sets. The statistical information of estimation errors of each subsystem state is illustrated in Figure~\ref{fig:duffing_1}. 

Both proposed approaches, mEDMD and mgEDMD, outperform other existing approaches due to a better performance in the second and third subsystem. All the four methods, i.e. sEDMD, lEDMD, mEDMD and mgEDMD, yield a similar performance in the first subsystem, which is easily explainable, considering the underlying graph structure. It can be observed that the error for $x_3$ is slightly larger than that for $x_2$. This difference could be attributed to the stronger impact of the estimation error of $x_1$ on  $x_3$ via the larger input gain $\gamma_3$. Furthermore,  the comparatively large  estimation errors of sEDMD  for $x_2$ and $x_3$ are hardly reduced by increasing the amount of training data while the improvement of lEDMD is observed.  This,  as previously discussed,  can be traced back to the inadequate state dimension of the predictors built for the extended subsystems corresponding to $\bar{x}_2$ and $\bar{x}_3$.

\begin{figure}[h!]
\centering

\tikzset{every picture/.style={line width=0.75pt}} %

\begin{tikzpicture}[x=0.75pt,y=0.75pt,yscale=-0.7,xscale=0.7]
\draw  [line width=1.5]  (151,135) .. controls (151,121.19) and (162.19,110) .. (176,110) .. controls (189.81,110) and (201,121.19) .. (201,135) .. controls (201,148.81) and (189.81,160) .. (176,160) .. controls (162.19,160) and (151,148.81) .. (151,135) -- cycle ;
\draw  [line width=1.5]  (94,210) .. controls (94,196.19) and (105.19,185) .. (119,185) .. controls (132.81,185) and (144,196.19) .. (144,210) .. controls (144,223.81) and (132.81,235) .. (119,235) .. controls (105.19,235) and (94,223.81) .. (94,210) -- cycle ;
\draw  [line width=1.5]  (208,210) .. controls (208,196.19) and (219.19,185) .. (233,185) .. controls (246.81,185) and (258,196.19) .. (258,210) .. controls (258,223.81) and (246.81,235) .. (233,235) .. controls (219.19,235) and (208,223.81) .. (208,210) -- cycle ;
\draw [line width=1.5]    (193,152.5) -- (217.64,186.27) ;
\draw [shift={(220,189.5)}, rotate = 233.88] [fill={rgb, 255:red, 0; green, 0; blue, 0 }  ][line width=0.08]  [draw opacity=0] (15.6,-3.9) -- (0,0) -- (15.6,3.9) -- cycle    ;
\draw [line width=1.5]    (159,153.75) -- (137.16,187.88) ;
\draw [shift={(135,191.25)}, rotate = 302.62] [fill={rgb, 255:red, 0; green, 0; blue, 0 }  ][line width=0.08]  [draw opacity=0] (15.6,-3.9) -- (0,0) -- (15.6,3.9) -- cycle    ;
\draw [line width=1.5]    (149,212.75) -- (203.5,212.75) ;
\draw [shift={(207.5,212.75)}, rotate = 180] [fill={rgb, 255:red, 0; green, 0; blue, 0 }  ][line width=0.08]  [draw opacity=0] (15.6,-3.9) -- (0,0) -- (15.6,3.9) -- cycle    ;
\draw [shift={(145,212.75)}, rotate = 0] [fill={rgb, 255:red, 0; green, 0; blue, 0 }  ][line width=0.08]  [draw opacity=0] (15.6,-3.9) -- (0,0) -- (15.6,3.9) -- cycle    ;

\draw (176.22,133.7) node  [font=\large]  {$1$};
\draw (118.5,208.5) node  [font=\large]  {$2$};
\draw (232.5,209) node  [font=\large]  {$3$};

\end{tikzpicture}

    \caption{Interconnected Van-der-Pol oscillators }
    \label{fig:dense_vandePol}
\end{figure}
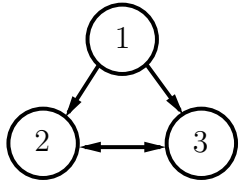

\noindent \textbf{System with output coupling: Van-der-Pol oscillators.} \label{CVO}  The coupled unforced Van-der-Pol oscillator adapted  from \cite{dutra2003modeling} is given by 
\begin{align*}
    \dot{x}_{1,1} &=  x_{1,2}, \ \dot{x}_{1,2} = 0.1 (1-\alpha_1 x_{1,1}^2 x_{1,2}) -\beta_1 x_{1,1},\\
     y_1  &= x_{1,1} x_{1,2}, \\
    \dot{x}_{i,1} &= x_{i,2}, \ \dot{x}_{i,2} = 0.01 (1- \alpha_i  x_{i,1} ^2x_{i,2})-\beta_i x_{i,1} +0.001 y_1 +0.1 (x_{i,2}- y_{j} ), \\
      y_i &= x_{i,2}, \qquad i \in \{2,3\}, j\in \{ 2,3 \} \setminus \{ i \},
\end{align*}
where $\alpha_1= 5.2525$, $\alpha_2=196.848$, $\alpha_3=5266.8$, $\beta_1=1$ and $\beta_2=\beta_3=4$. The underlying  interconnection is depicted in Figure~\ref{fig:dense_vandePol}.

 Two data sets consisting $\bar{m}=2500$ and $\bar{m}=8000$ pairs of snapshots, which are collected in single time-step with $\Delta_t=0.01s$,  are used to construct predictors. The initial conditions for each data set are generated randomly on $\X=([-\pi/2, \pi/2] \times [-1,1])^3$. 
 To construct data matrices for each subsystem,    $m_i=\bar{m}$ pairs of samples  extracted from each data set are used. 
The overall number of observables is $N=456$. The configuration of all implemented approaches are listed below 
\begin{itemize}
  \item \emph{EDMD:}  The observable  is $\bar{\varphi}(x) = [x^\top \  \varphi^1(x) \ \ldots \ \varphi^{N-6}(x)]^\top$ with $x=[x_1^\top \ x_2^\top \ x_3^\top]^\top$ and thin plate RBFs  in \eqref{eq:RBF}  $\varphi^k$  with centers randomly selected over $\X$ for all $k$. 
    \item \emph{lEDMD:} The observable is chosen to be  $\bar{\varphi}_i (x_i) =  [x_i^\top \  \varphi_i^1(x_i) \ \ldots \ \varphi_i^{N_i-2}(x_i)]^\top$ for each subsystem indexed by $i \in [1:3]$, where $N_i = N/3$,  $\varphi^k_i$ is a thin plate RBF with a center randomly chosen in $\X_i= [-\pi/2, \pi/2] \times [-1,1]$ for each $k$.  
 \item \emph{mEDMD:}  For the $1$-st subsystem and $i$-th subsystem with $i\in \{2,3\}$, their observables are  $\bar{\varphi}_1 (x_1) =  [x_1^\top \ y_1 \ \varphi_1^1(x_1) \ \ldots \ \varphi_1^{N_1-3}(x_1)]^\top$ and $\bar{\varphi}_i (x_i) =  [x_i^\top \  \varphi_i^1(x_i) \ \ldots \ \varphi_i^{N_i-2}(x_i)]^\top$ respectively, where  $\varphi_i^k$ is a  thin plate RBF with a center randomly chosen in $\X_i$. 
\end{itemize}

The $2$nd or the $3$rd subsystem together with its in-neighbours  constitute the vertices set of the graph. Hence, building a predictor for the $2$nd or the $3$rd subsystem with sparse EDMD reduces to the predictor construction by EDMD. Therefore, sEDMD is not implemented. As the graph of the whole system involves one  cycle, we are in the setting of   Proposition~\ref{theo:case_2} meaning that the condition~\eqref{eq:smallgain} is applicable.

\begin{figure}[h]
\centering
\vspace{-10pt}
\begin{subfigure}{0.4\textwidth}
\hspace{75pt}
  \includegraphics[width=0.28\textwidth, clip=false, trim = 60mm 100mm 55mm 62mm]{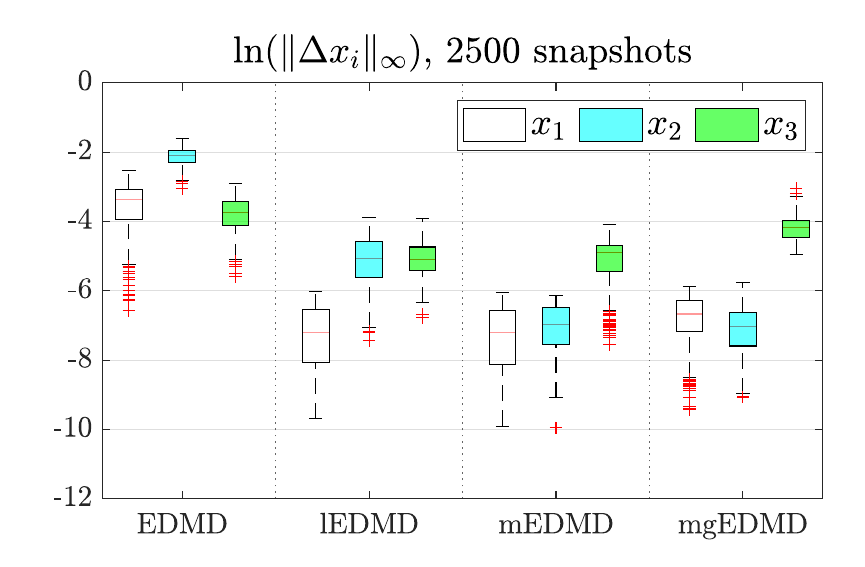}
\end{subfigure}
\hfill
\begin{subfigure}{0.4\textwidth}
\hspace{45pt}
\includegraphics[width=0.28\textwidth, clip=false, trim = 60mm 100mm 55mm 62mm]{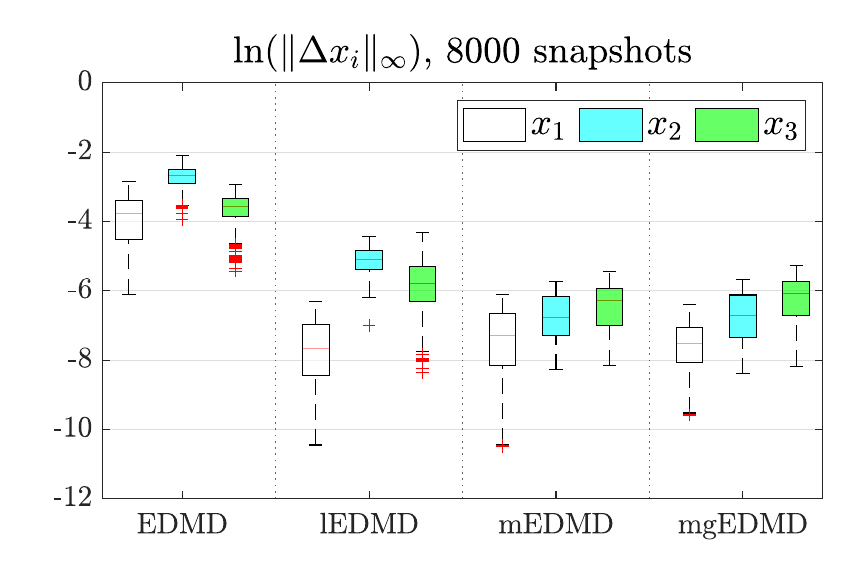}
\end{subfigure}
\vspace{150pt}
\caption{Prediction errors for Van-der-Pol oscillators }
\label{fig:Pol_1}
\end{figure}

For the evaluation, we implement $500$ simulations with randomly chosen initial conditions in $([-\pi/5, \pi/5]\times [-2/5, 2/5])^3$. Similar to  the Duffing system,  the trajectory is predicted over the time interval $[0, 0.5s]$  and  $\ln(\|\Delta x_i\|_\infty)$  for each subsystem is computed in each simulation. The statistical information of all the obtained  errors is  illustrated in Figure~\ref{fig:Pol_1}.  The estimation performance of the  proposed mEDMD and mgEDMD is better than that of  sEDMD and lEDMD in $x_2$, while lEDMD, mEDMD and mgEDMD yield similiar estimation errors in $x_1$ due to the underlying graph structure. The  condition~\eqref{eq:smallgain} mentioned previously is verified to be unsatisfied. Hence,  Proposition~\ref{theo:case_2} can not be employed to deduce error bounds.  Nevertheless, as shown in Figure~\ref{fig:Pol_1}, the errors of mgEDMD tend to decrease as more data is utilized for learning,  implying the potential to relax  the condition~\eqref{eq:smallgain} for error analysis.

\section{Data-efficient transfer learning via modularized EDMD}\label{sec:transfer}

\noindent 
To demonstrate  the capability of the proposed modularized EDMD approach in terms of transfer learning, %
we  compare  prediction errors of  lEDMD as introduced Subsection~\ref{Overview} and the proposed approaches by means of three examples involving coupled Duffing oscillators. As the predictors of both EDMD and sEDMD in Subsection~\ref{Overview} are built on global or extended systems, they do not enable efficient transfer learning and are thus not addressed in this section.    

\subsection{Transfer Learning between identical subsystems.}
\begin{figure}[h]
    \centering
 \tikzset{every picture/.style={line width=0.75pt}} %

\begin{tikzpicture}[x=0.75pt,y=0.75pt,yscale=-0.7,xscale=0.7]
\draw  [line width=1.5]  (151,135) .. controls (151,121.19) and (162.19,110) .. (176,110) .. controls (189.81,110) and (201,121.19) .. (201,135) .. controls (201,148.81) and (189.81,160) .. (176,160) .. controls (162.19,160) and (151,148.81) .. (151,135) -- cycle ;
\draw  [line width=1.5]  (94,210) .. controls (94,196.19) and (105.19,185) .. (119,185) .. controls (132.81,185) and (144,196.19) .. (144,210) .. controls (144,223.81) and (132.81,235) .. (119,235) .. controls (105.19,235) and (94,223.81) .. (94,210) -- cycle ;
\draw  [line width=1.5]  (208,210) .. controls (208,196.19) and (219.19,185) .. (233,185) .. controls (246.81,185) and (258,196.19) .. (258,210) .. controls (258,223.81) and (246.81,235) .. (233,235) .. controls (219.19,235) and (208,223.81) .. (208,210) -- cycle ;
\draw [line width=1.5]    (193,152.5) -- (217.64,186.27) ;
\draw [shift={(220,189.5)}, rotate = 233.88] [fill={rgb, 255:red, 0; green, 0; blue, 0 }  ][line width=0.08]  [draw opacity=0] (15.6,-3.9) -- (0,0) -- (15.6,3.9) -- cycle    ;
\draw [line width=1.5]    (159,153.75) -- (137.16,187.88) ;
\draw [shift={(135,191.25)}, rotate = 302.62] [fill={rgb, 255:red, 0; green, 0; blue, 0 }  ][line width=0.08]  [draw opacity=0] (15.6,-3.9) -- (0,0) -- (15.6,3.9) -- cycle    ;
\draw  [fill={rgb, 255:red, 0; green, 0; blue, 0 }  ,fill opacity=1 ] (286,168.92) -- (303.8,168.92) -- (303.8,165) -- (315.67,172.83) -- (303.8,180.67) -- (303.8,176.75) -- (286,176.75) -- cycle ;
\draw  [line width=1.5]  (405.67,135.67) .. controls (405.67,121.86) and (416.86,110.67) .. (430.67,110.67) .. controls (444.47,110.67) and (455.67,121.86) .. (455.67,135.67) .. controls (455.67,149.47) and (444.47,160.67) .. (430.67,160.67) .. controls (416.86,160.67) and (405.67,149.47) .. (405.67,135.67) -- cycle ;
\draw  [line width=1.5]  (348.67,210.67) .. controls (348.67,196.86) and (359.86,185.67) .. (373.67,185.67) .. controls (387.47,185.67) and (398.67,196.86) .. (398.67,210.67) .. controls (398.67,224.47) and (387.47,235.67) .. (373.67,235.67) .. controls (359.86,235.67) and (348.67,224.47) .. (348.67,210.67) -- cycle ;
\draw  [line width=1.5]  (462.67,210.67) .. controls (462.67,196.86) and (473.86,185.67) .. (487.67,185.67) .. controls (501.47,185.67) and (512.67,196.86) .. (512.67,210.67) .. controls (512.67,224.47) and (501.47,235.67) .. (487.67,235.67) .. controls (473.86,235.67) and (462.67,224.47) .. (462.67,210.67) -- cycle ;
\draw [line width=1.5]    (447.67,153.17) -- (472.31,186.94) ;
\draw [shift={(474.67,190.17)}, rotate = 233.88] [fill={rgb, 255:red, 0; green, 0; blue, 0 }  ][line width=0.08]  [draw opacity=0] (15.6,-3.9) -- (0,0) -- (15.6,3.9) -- cycle    ;
\draw [line width=1.5]    (413.67,154.42) -- (391.82,188.55) ;
\draw [shift={(389.67,191.92)}, rotate = 302.62] [fill={rgb, 255:red, 0; green, 0; blue, 0 }  ][line width=0.08]  [draw opacity=0] (15.6,-3.9) -- (0,0) -- (15.6,3.9) -- cycle    ;
\draw [line width=1.5]    (512.67,210.68) -- (570.67,210.82) ;
\draw [shift={(574.67,210.83)}, rotate = 180.15] [fill={rgb, 255:red, 0; green, 0; blue, 0 }  ][line width=0.08]  [draw opacity=0] (15.6,-3.9) -- (0,0) -- (15.6,3.9) -- cycle    ;
\draw  [line width=1.5]  (575,210) .. controls (575,196.19) and (586.19,185) .. (600,185) .. controls (613.81,185) and (625,196.19) .. (625,210) .. controls (625,223.81) and (613.81,235) .. (600,235) .. controls (586.19,235) and (575,223.81) .. (575,210) -- cycle ;

\draw (176.22,133.7) node  [font=\large]  {$1$};
\draw (118.5,208.5) node  [font=\large]  {$2$};
\draw (232.5,209) node  [font=\large]  {$3$};
\draw (430.89,134.36) node  [font=\large]  {$1$};
\draw (373.17,209.17) node  [font=\large]  {$2$};
\draw (487.17,209.67) node  [font=\large]  {$3$};
\draw (599.5,208.5) node  [font=\large]  {$4$};
\end{tikzpicture}
    \caption{Transfer learning within identical subsystems }
    \label{fig:Graph_transfer_1}
\end{figure}
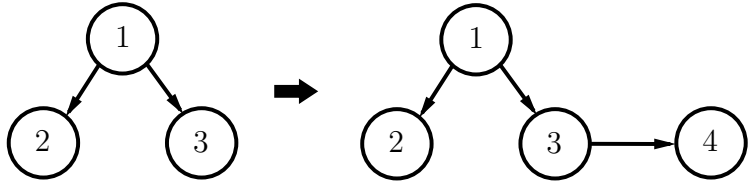

\begin{figure}[h]
\centering
\vspace{-10pt}
\begin{subfigure}{0.4\textwidth}
\hspace{77pt}
  \includegraphics[width=0.29\textwidth, clip=false, trim = 60mm 100mm 55mm 62mm]{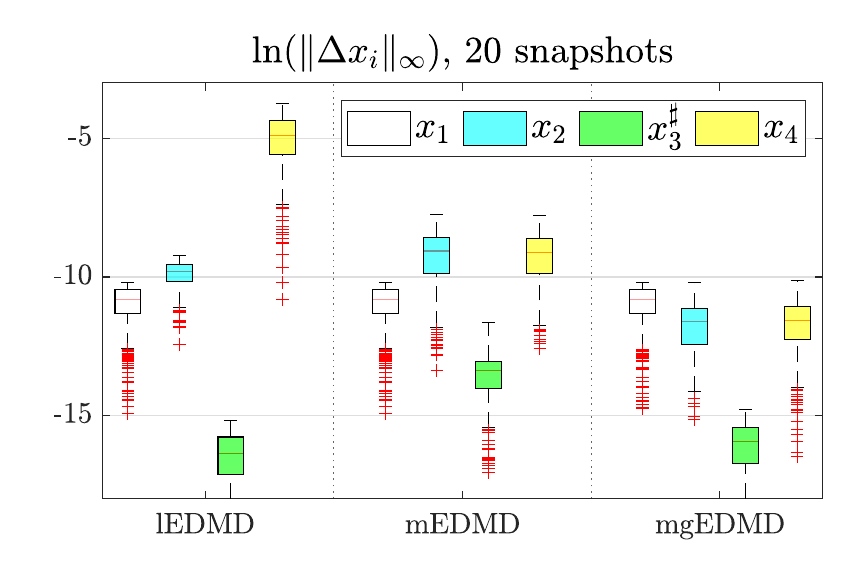}
\end{subfigure}
\hfill
\begin{subfigure}{0.4\textwidth}
\hspace{45pt}
\includegraphics[width=0.29\textwidth, clip=false, trim = 60mm 100mm 55mm 62mm]{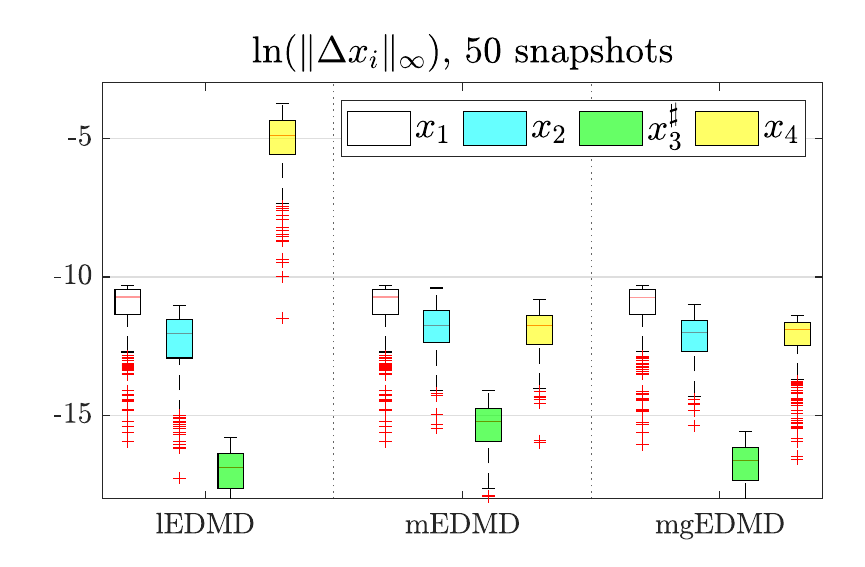}
\end{subfigure}
\vspace{155pt}
\caption{Errors for Duffing oscillators for Figure~\ref{fig:Graph_transfer_1}}
\label{fig:Duff_transfer_ga30}
\end{figure}

 Let us  consider  the following coupled Duffing oscillators,
\begin{subequations} \label{eq:example_transfer}
   \begin{align}
    \dot{x}_{1,1} &= \alpha_1 x_{1,2}, \  \dot{x}_{1,2}  = -\beta_1 x_{1,1}^3, \\
    \dot{x}_{i,1} &= \alpha_i x_{i,2}, \ \dot{x}_{i,2}  = -\beta_i x_{i,1}^3 + \gamma_i x_{1,1},  \qquad  i\in \{2,3\},  
    \end{align} 
\end{subequations}
with  $(\alpha_1, \alpha_2, \alpha_3, \beta_1,\beta_2, \beta_3, \gamma_2, \gamma_3) = ( 0.2, 0.06, 0.004, 0.1, 0.08, 0.03, 0.05, 0.001)$. 
We choose monomials with order up to three as observables for each subsystem and then employ the lEDMD, mEDMD and mgEDMD to identify the system. For learning, two data sets consisting $20$ and $50$ pairs of snapshots are collected with time-step of $\Delta_t = 0.01$, whereby initial conditions  are chosen randomly on the box $\X = [-1.5, 1.5]^6$. After building  predictors by the different approaches, we modify the system by adding a subsystem labelled by $4$, as depicted in Figure~\ref{fig:Graph_transfer_1}. 
The $4$th subsystem exhibits the same dynamic as $2$nd subsystem, henceforth, the  predictor associated with the $2$nd subsystem is carried over to $4$th subsystem. For the evaluation, 500 simulations are implemented  with 50 prediction time steps and randomly chosen initial conditions in $ [-0.5, 0.5]^8$.  The prediction errors $\ln{ \infn{\Delta x_i}  }$ by different approaches are   statistically shown in Figure~\ref{fig:Duff_transfer_ga30} for each subsystem.  Since the  graph is still acyclic after modification,  the error bound of  Proposition~\ref{prop:acyclic} applies. The comparable significant error of $x_4$ for lEDMD demonstrates  the ability of  our approach in terms of  transfer learning among identical subsystems, i.e $2$nd and $4$th subsystems, especially when more data are available.  

\subsection{Learning between partially identical subsystems.} 
\begin{figure}[h]
    \centering
 \tikzset{every picture/.style={line width=0.75pt}} %

\begin{tikzpicture}[x=0.75pt,y=0.75pt,yscale=-0.7,xscale=0.7]
\draw  [line width=1.5]  (151,135) .. controls (151,121.19) and (162.19,110) .. (176,110) .. controls (189.81,110) and (201,121.19) .. (201,135) .. controls (201,148.81) and (189.81,160) .. (176,160) .. controls (162.19,160) and (151,148.81) .. (151,135) -- cycle ;
\draw  [line width=1.5]  (94,210) .. controls (94,196.19) and (105.19,185) .. (119,185) .. controls (132.81,185) and (144,196.19) .. (144,210) .. controls (144,223.81) and (132.81,235) .. (119,235) .. controls (105.19,235) and (94,223.81) .. (94,210) -- cycle ;
\draw  [line width=1.5]  (208,210) .. controls (208,196.19) and (219.19,185) .. (233,185) .. controls (246.81,185) and (258,196.19) .. (258,210) .. controls (258,223.81) and (246.81,235) .. (233,235) .. controls (219.19,235) and (208,223.81) .. (208,210) -- cycle ;
\draw [line width=1.5]    (193,152.5) -- (217.64,186.27) ;
\draw [shift={(220,189.5)}, rotate = 233.88] [fill={rgb, 255:red, 0; green, 0; blue, 0 }  ][line width=0.08]  [draw opacity=0] (15.6,-3.9) -- (0,0) -- (15.6,3.9) -- cycle    ;
\draw [line width=1.5]    (159,153.75) -- (137.16,187.88) ;
\draw [shift={(135,191.25)}, rotate = 302.62] [fill={rgb, 255:red, 0; green, 0; blue, 0 }  ][line width=0.08]  [draw opacity=0] (15.6,-3.9) -- (0,0) -- (15.6,3.9) -- cycle    ;
\draw  [fill={rgb, 255:red, 0; green, 0; blue, 0 }  ,fill opacity=1 ] (286,168.92) -- (303.8,168.92) -- (303.8,165) -- (315.67,172.83) -- (303.8,180.67) -- (303.8,176.75) -- (286,176.75) -- cycle ;
\draw  [line width=1.5]  (405.67,135.67) .. controls (405.67,121.86) and (416.86,110.67) .. (430.67,110.67) .. controls (444.47,110.67) and (455.67,121.86) .. (455.67,135.67) .. controls (455.67,149.47) and (444.47,160.67) .. (430.67,160.67) .. controls (416.86,160.67) and (405.67,149.47) .. (405.67,135.67) -- cycle ;
\draw  [line width=1.5]  (348.67,210.67) .. controls (348.67,196.86) and (359.86,185.67) .. (373.67,185.67) .. controls (387.47,185.67) and (398.67,196.86) .. (398.67,210.67) .. controls (398.67,224.47) and (387.47,235.67) .. (373.67,235.67) .. controls (359.86,235.67) and (348.67,224.47) .. (348.67,210.67) -- cycle ;
\draw  [line width=1.5]  (462.67,210.67) .. controls (462.67,196.86) and (473.86,185.67) .. (487.67,185.67) .. controls (501.47,185.67) and (512.67,196.86) .. (512.67,210.67) .. controls (512.67,224.47) and (501.47,235.67) .. (487.67,235.67) .. controls (473.86,235.67) and (462.67,224.47) .. (462.67,210.67) -- cycle ;
\draw [line width=1.5]    (447.67,153.17) -- (472.31,186.94) ;
\draw [shift={(474.67,190.17)}, rotate = 233.88] [fill={rgb, 255:red, 0; green, 0; blue, 0 }  ][line width=0.08]  [draw opacity=0] (15.6,-3.9) -- (0,0) -- (15.6,3.9) -- cycle    ;
\draw [line width=1.5]    (413.67,154.42) -- (391.82,188.55) ;
\draw [shift={(389.67,191.92)}, rotate = 302.62] [fill={rgb, 255:red, 0; green, 0; blue, 0 }  ][line width=0.08]  [draw opacity=0] (15.6,-3.9) -- (0,0) -- (15.6,3.9) -- cycle    ;
\draw [line width=1.5]    (400,210.75) -- (460.5,210.75) ;
\draw [shift={(460.5,210.75)}, rotate = 180] [fill={rgb, 255:red, 0; green, 0; blue, 0 }  ][line width=0.08]  [draw opacity=0] (15.6,-3.9) -- (0,0) -- (15.6,3.9) -- cycle    ;
\draw (176.22,133.7) node  [font=\large]  {$1$};
\draw (118.5,208.5) node  [font=\large]  {$2$};
\draw (232.5,209) node  [font=\large]  {$3$};
\draw (430.89,134.36) node  [font=\large]  {$1$};
\draw (373.17,209.17) node  [font=\large]  {$2$};
\draw (487.17,209.67) node  [font=\large]  {$3^{\sharp}$};

\end{tikzpicture}
    \caption{Learning within partially identical subsystems}
    \label{fig:Graph_transfer_2}
\end{figure}
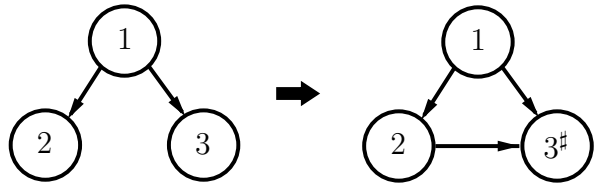

Let us consider the original system~\eqref{eq:example_transfer} and modify it by adding an interaction to $3$rd subsystem and label the modified $3$rd subsystem by $3^\sharp$, whose dynamic is described by 
\begin{align*}
     \dot{x}_{3^\sharp,1} &= \alpha_3 x_{3^\sharp,2},  \ \dot{x}_{3^\sharp,2}  = -\beta_3 x_{3^\sharp,1}^3 + \gamma_3 x_{1,1} + 0.08 x_{3^\sharp,2} x_{2,2},\\ 
\end{align*}
The modification is illustrated in Figure~\ref{fig:Graph_transfer_2}. 

\begin{figure}[h]
\centering
\vspace{-10pt}
\begin{subfigure}{0.4\textwidth}
\hspace{77pt}
  \includegraphics[width=0.29\textwidth, clip=false, trim = 60mm 100mm 55mm 62mm]{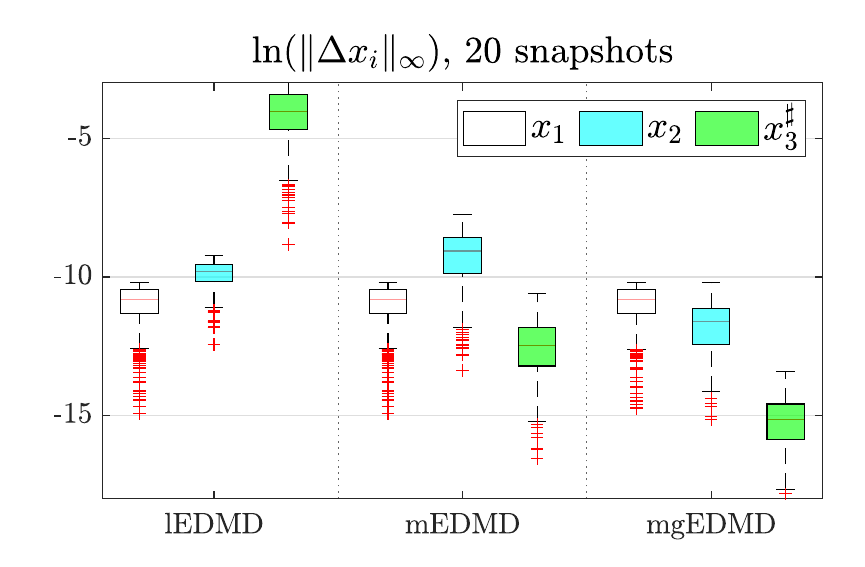}
\end{subfigure}
\hfill
\begin{subfigure}{0.4\textwidth}
\hspace{45pt}
\includegraphics[width=0.29\textwidth, clip=false, trim = 60mm 100mm 55mm 62mm]{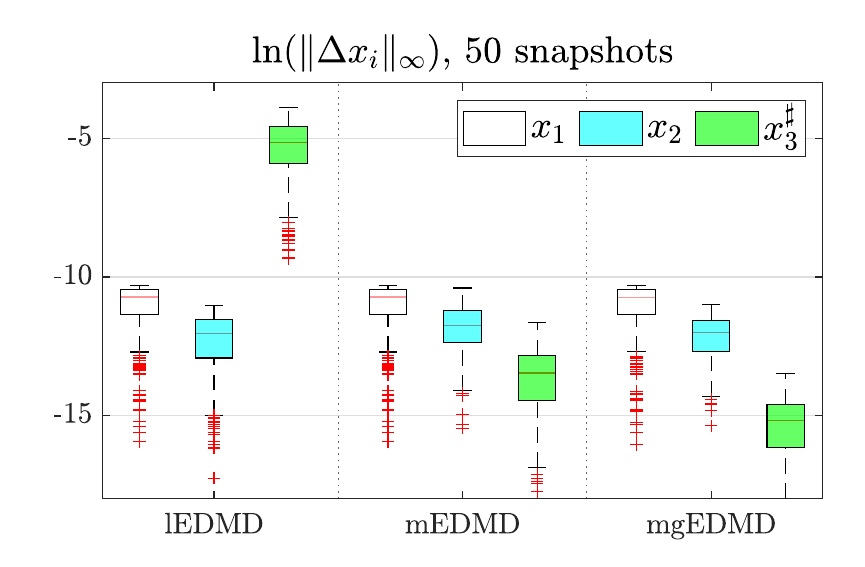}
\end{subfigure}
\vspace{155pt}
\caption{Errors for Duffing oscillators for Figure~\ref{fig:Graph_transfer_2}}
\label{fig:Duff_transfer_ga3_N_3}
\end{figure}

Drawing upon the predictors built for the original system~\eqref{eq:example_transfer}, only the input matrix $B_{3^\sharp,2}$ in \eqref{eq:lDMD} for lEDMD  and the matrix $K_{3^\sharp,2}^{\Delta_t}$ ($(L_{m_{3^\sharp}}^{e^r_2})_{r\in[1:2]}$ ) in~\eqref{eq:mEDMD_operator} (Algorithm~\ref{Algo: mgEDMD}) for gEDMD (mgEDMD)  are computed  from  newly randomly generated $20$ and $50$ pairs of snapshots over $\X$. The configuration for evaluation is similar as above, except that initial conditions are selected within $[-0.5, 0.5]^6$ for simulations.  As, again, the underlying graph is acyclic, Proposition~\ref{prop:acyclic}  is applicable to derive estimation errors for mgEDMD.   The prediction errors by employing all approaches are shown in the Figure~\ref{fig:Duff_transfer_ga3_N_3}, which clearly show the effectiveness of the proposed approach in  transfer learning between the $3$rd subsystem and $3^\sharp$rd subsystem.   

\subsection{Learning within identical and partially identical subsystems.} In the last case,  we modify the orginal system~\eqref{eq:example_transfer} by adding an input channel into $3$rd subsystem and  incorporating a new subsystem labeled as $4$. The  modified $3$rd subsystem is labeled by $3^\sharp$.  The  $4$th subsystem, which has an identical dynamic as the $2$nd subsystem,   interacts with the $3^\sharp$rd subsystem as follows,
\begin{align*}
     \dot{x}_{3^\sharp,1} &= \alpha_3 x_{3,2},  \ \dot{x}_{3^\sharp,2}  = -\beta_3 x_{3^\sharp,1}^3 + \gamma_3 x_{1,1} + 0.08 x_{3^\sharp,2} x_{4,2},\\
     \dot{x}_{4,1} &= \alpha_2 x_{4,2}, \ \dot{x}_{4,2}  = -\beta_2 x_{4,1}^3 + \gamma_2 x_{3^\sharp,2}. 
\end{align*}

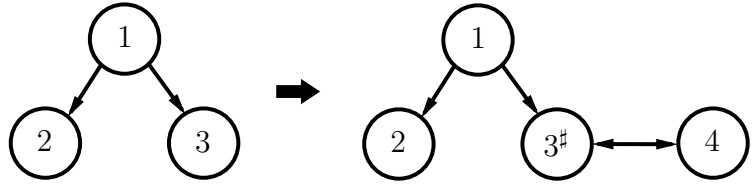
\begin{figure}[h]
    \centering
 \tikzset{every picture/.style={line width=0.75pt}} %

\begin{tikzpicture}[x=0.75pt,y=0.75pt,yscale=-0.7,xscale=0.7]
\draw  [line width=1.5]  (151,135) .. controls (151,121.19) and (162.19,110) .. (176,110) .. controls (189.81,110) and (201,121.19) .. (201,135) .. controls (201,148.81) and (189.81,160) .. (176,160) .. controls (162.19,160) and (151,148.81) .. (151,135) -- cycle ;
\draw  [line width=1.5]  (94,210) .. controls (94,196.19) and (105.19,185) .. (119,185) .. controls (132.81,185) and (144,196.19) .. (144,210) .. controls (144,223.81) and (132.81,235) .. (119,235) .. controls (105.19,235) and (94,223.81) .. (94,210) -- cycle ;
\draw  [line width=1.5]  (208,210) .. controls (208,196.19) and (219.19,185) .. (233,185) .. controls (246.81,185) and (258,196.19) .. (258,210) .. controls (258,223.81) and (246.81,235) .. (233,235) .. controls (219.19,235) and (208,223.81) .. (208,210) -- cycle ;
\draw [line width=1.5]    (193,152.5) -- (217.64,186.27) ;
\draw [shift={(220,189.5)}, rotate = 233.88] [fill={rgb, 255:red, 0; green, 0; blue, 0 }  ][line width=0.08]  [draw opacity=0] (15.6,-3.9) -- (0,0) -- (15.6,3.9) -- cycle    ;
\draw [line width=1.5]    (159,153.75) -- (137.16,187.88) ;
\draw [shift={(135,191.25)}, rotate = 302.62] [fill={rgb, 255:red, 0; green, 0; blue, 0 }  ][line width=0.08]  [draw opacity=0] (15.6,-3.9) -- (0,0) -- (15.6,3.9) -- cycle    ;
\draw  [fill={rgb, 255:red, 0; green, 0; blue, 0 }  ,fill opacity=1 ] (286,168.92) -- (303.8,168.92) -- (303.8,165) -- (315.67,172.83) -- (303.8,180.67) -- (303.8,176.75) -- (286,176.75) -- cycle ;
\draw  [line width=1.5]  (405.67,135.67) .. controls (405.67,121.86) and (416.86,110.67) .. (430.67,110.67) .. controls (444.47,110.67) and (455.67,121.86) .. (455.67,135.67) .. controls (455.67,149.47) and (444.47,160.67) .. (430.67,160.67) .. controls (416.86,160.67) and (405.67,149.47) .. (405.67,135.67) -- cycle ;
\draw  [line width=1.5]  (348.67,210.67) .. controls (348.67,196.86) and (359.86,185.67) .. (373.67,185.67) .. controls (387.47,185.67) and (398.67,196.86) .. (398.67,210.67) .. controls (398.67,224.47) and (387.47,235.67) .. (373.67,235.67) .. controls (359.86,235.67) and (348.67,224.47) .. (348.67,210.67) -- cycle ;
\draw  [line width=1.5]  (462.67,210.67) .. controls (462.67,196.86) and (473.86,185.67) .. (487.67,185.67) .. controls (501.47,185.67) and (512.67,196.86) .. (512.67,210.67) .. controls (512.67,224.47) and (501.47,235.67) .. (487.67,235.67) .. controls (473.86,235.67) and (462.67,224.47) .. (462.67,210.67) -- cycle ;
\draw [line width=1.5]    (447.67,153.17) -- (472.31,186.94) ;
\draw [shift={(474.67,190.17)}, rotate = 233.88] [fill={rgb, 255:red, 0; green, 0; blue, 0 }  ][line width=0.08]  [draw opacity=0] (15.6,-3.9) -- (0,0) -- (15.6,3.9) -- cycle    ;
\draw [line width=1.5]    (413.67,154.42) -- (391.82,188.55) ;
\draw [shift={(389.67,191.92)}, rotate = 302.62] [fill={rgb, 255:red, 0; green, 0; blue, 0 }  ][line width=0.08]  [draw opacity=0] (15.6,-3.9) -- (0,0) -- (15.6,3.9) -- cycle    ;
\draw [line width=1.5]    (516.67,210.68) -- (570.67,210.82) ;
\draw [shift={(574.67,210.83)}, rotate = 180.15] [fill={rgb, 255:red, 0; green, 0; blue, 0 }  ][line width=0.08]  [draw opacity=0] (15.6,-3.9) -- (0,0) -- (15.6,3.9) -- cycle    ;
\draw [shift={(512.67,210.67)}, rotate = 0.15] [fill={rgb, 255:red, 0; green, 0; blue, 0 }  ][line width=0.08]  [draw opacity=0] (15.6,-3.9) -- (0,0) -- (15.6,3.9) -- cycle    ;
\draw  [line width=1.5]  (575,210) .. controls (575,196.19) and (586.19,185) .. (600,185) .. controls (613.81,185) and (625,196.19) .. (625,210) .. controls (625,223.81) and (613.81,235) .. (600,235) .. controls (586.19,235) and (575,223.81) .. (575,210) -- cycle ;

\draw (176.22,133.7) node  [font=\large]  {$1$};
\draw (118.5,208.5) node  [font=\large]  {$2$};
\draw (232.5,209) node  [font=\large]  {$3$};
\draw (430.89,134.36) node  [font=\large]  {$1$};
\draw (373.17,209.17) node  [font=\large]  {$2$};
\draw (487.17,209.67) node  [font=\large]  {$3^\sharp$};
\draw (599.5,208.5) node  [font=\large]  {$4$};

\end{tikzpicture}
    \caption{Learning within identical and partially identical subsystems }
    \label{fig:Graph_transfer}
\end{figure}

\noindent The modification in terms of interconnections  is illustrated in the Figure~\ref{fig:Graph_transfer}.

 For lEDMD and our approaches, the  predictor built for the $2$nd subsystem is also the predictor for the $4$th subsystem.  Furthermore, for mEDMD (mgEDMD),  only the new matrix $K_{3^\sharp,4}^{\Delta_t}$ ($(L_{m_{3^\sharp}}^{e^r_4})_{r\in[1:2]}$ ) associated to the coupling term $x_4$  in \eqref{eq:mEDMD_operator} (Algorithm~\ref{Algo: mgEDMD}) is identified, while the rest  are inherited from  the  predictor of the original $3$rd subsystem. 
Since the new graph contains a cycle, the condition~\eqref{eq:smallgain} in Proposition~\ref{theo:case_2} is applicable with  $E_{3^\sharp4}E_{43^\sharp}= 0.803 < 1$, whereby $E_{3^\sharp4}$ and $E_{43^\sharp}$ are approximately computed based on $2000$ snapshots. Hence, Proposition~\ref{theo:case_2} can be employed to analysis error bounds for mgEDMD.   As for the lEDMD approach,   only  the input matrix $B_{3^\sharp,4}$ in \eqref{eq:lDMD} is identified  and all the other model parameters are carried over  from the constructed predictor for the orignal $3$rd subsystem. The setting for  evaluation is the same as the example corresponding to  Figure~\ref{fig:Graph_transfer_1}. 

\begin{figure}[ht]
\centering
\vspace{-10pt}
\begin{subfigure}{0.4\textwidth}
\hspace{77pt}
  \includegraphics[width=0.29\textwidth, clip=false, trim = 60mm 100mm 55mm 62mm]{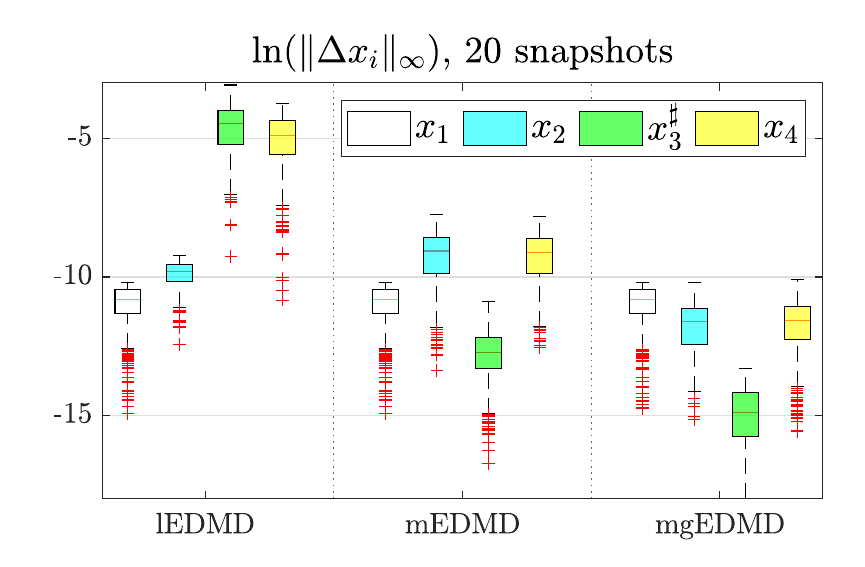}
\end{subfigure}
\hfill
\begin{subfigure}{0.4\textwidth}
\hspace{45pt}
\includegraphics[width=0.29\textwidth, clip=false, trim = 60mm 100mm 55mm 62mm]{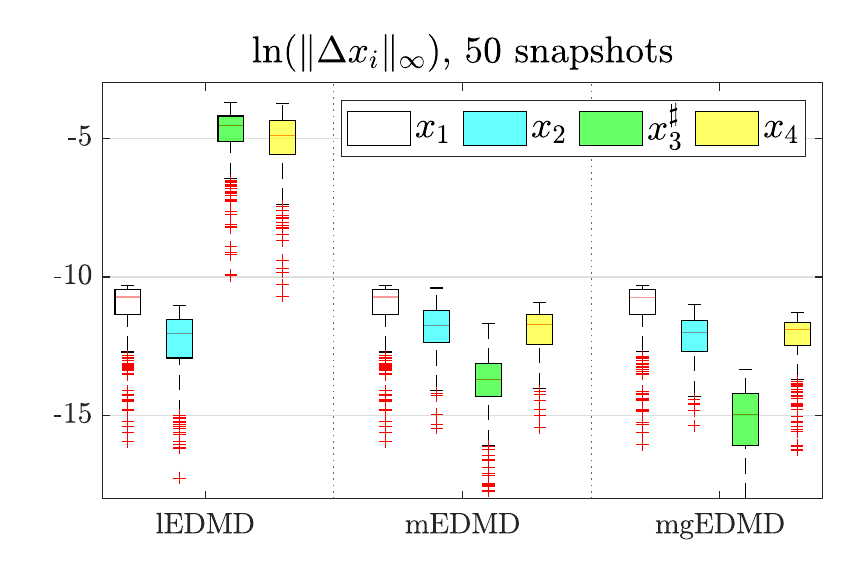}
\end{subfigure}
\vspace{160pt}
\caption{Errors for  Duffing oscillators for Figure~\ref{fig:Graph_transfer} }
\label{fig:Duff_transfer}
\end{figure}

The results in Figure~\ref{fig:Duff_transfer} clearly illustrates the advantage of our proposed approach.  The comparably huge errors of \nametbdd{lEDMD} in this example and also in the other two examples  trace back to the fact that, the interactions within  original subsystems can not be captured accurately by local linear Koopman models. Specifically, by incorporating neighboring states into the subspace of observables  when constructing the linear Koopman model,  the local subsystem and its   in-neighbours  are  interconnected to each other and  collectively  treated  as a \emph{self-contained} system decoupled from all other systems in the graph.  Consequently, significant modelling errors can arise when digraph is changed, such as  by adding new subsystem as in-neighbours, which would necessitate a complete re-identification of some subsystems. %

\section{Conclusion and outlook} \label{sec:conclusion}
We proposed a modularized EDMD approach to learn Koopman models of interconnected systems. There, we leveraged the structure of the generator to formulate a learning architecture allowing to identify local models which are then combined to a global model. Under various different assumptions on the underlying topology of the interconnection, we proved finite-data estimation error bounds for the proposed method. Moreover, we showed that our approach leads to smaller prediction errors than related methods and facilitates transfer learning, thereby enhancing the efficiency of modelling in case of topology changes.  All the simulations are conducted in Matlab and available at \url{https://github.com/memopanda/KoopmanGraph.git}.

Future work considers the extension of the proposed EDMD variant to  systems with control inputs, e.g., for data-driven distributed control of large-scale nonlinear systems, and extend the estimation errors analysis to Koopman operators.  Moreover, exploring the identification of coupling structures and providing (possibly) sparse representation of the Koopman model for the whole system will be addressed in the future.   %

\bibliographystyle{abbrv}
\bibliography{references}

\appendix

\section{Auxiliary results}
\begin{lem} \label{Appendix:2}
  For a cyclic digraph having no vertex shared by cycles, its strong components are either single cycles or vertices.    
\end{lem}
\begin{proof}
    Let us assume that one strong component $\calI_k$ consists of multiple cycles, where no vertex lies in at least two cycles. Let $v_i$  be one vertex in once cycle  $G_1$ in $\calI_k$,  and $v_j$ be another vertex in another cycle $G_2$ in $\calI_k$. Due to the definition of strong component, $v_i$ and $v_j$ must be reached from each other, henceforth constitute a circle which shares $v_i$ and $v_j$ with cycles $G_1$ and $G_2$ respectively. This however contradicts the assumption. 
\end{proof}

\begin{lem} \label{Appendix:1}
Let $b=[b_1 \ \ldots \ b_n]^\top \in \R^{n}$, $n \in \mathbb{N}$,  be a positive vector, i.e\ $b_i>0$ for all $i\in [1:n]$. For $ A \in \R^{n\times n}$,   the entry in its $i$-th row and $j$-th column is denoted by $a_{ij}$.   Given  $\alpha  \subset [1:n]$ and $ \beta \subset [1:n]$,   let us denote by  $A_{\alpha,\beta}$  the  submatrix  of entries that lie in the rows of $A$ indexed by $\alpha$ and the columns indexed by $\beta$.  
\noindent Let us define 
 \begin{equation*}
     U^0=A, U^k=U^{k-1}_{[2:n], [2:n]}-(U^{k-1}_{[1], [1]})^{-1}U^{k-1}_{[2:n], [1]} U^{k-1}_{[1], [2:n]}, k=[1:n-1],   
 \end{equation*}
and then  a set $\mathcal{M}_n:=\{ A \in \R^{n\times n}: a_{ii}>0, \ a_{ij}\leq 0 \ \text{for} \ i\neq j, \ U^{k}_{[1],[1]} > 0, \forall k\in [1:n-1]\}$.

 Then, for any matrix  $A \in \mathcal{M}_n$, 
 $A$ is invertible and the non-empty set $\{ e\in \R^n_{\geq 0} : Ae\preceq b \} $ lies in $ \{ e\in \R^{n}_{\geq 0}: e\preceq A^{-1}b\}$ with $b \succeq \mathbf{0}$. 
 
 Furthermore, for $A, \widehat{A} \in \mathcal{M}_n$, 
 \begin{equation} \label{Appendix:monotone}
  0 \leq  -a_{ij} \leq -\widehat{a}_{ij} \ \forall i\neq j, \ a_{ii}=\widehat{a}_{ii}, \   0 < b_i \leq \widehat{b}_i \ \forall i \ \Rightarrow A^{-1} b \preceq \widehat{A}^{-1}\widehat{b}.  
 \end{equation}
 \end{lem}
\begin{proof}
 Define 
  \begin{equation*}
      L^{k}=\mas{ccc}{I_{k-1} & O & O \\ O & 1 & O \\ O & -U^{k-1}_{[2:n][1]} (U^{k-1}_{[1], [1]})^{-1} & I_{n-k} }, k \in [1 : n-1]
  \end{equation*}
After simple computation, we get
\begin{equation*}
   \left( \prod_{k=1}^{n-1} L^k  \right )A = 
   \mas{cccc}{ U^0_{[1],[1]} & U^0_{[1],[2]} & \dots & U^0_{[1],[n]}\\
               0  &  U^1_{[1],[1]} & \dots & U^1_{[1],[n]}  \\
               \vdots &    &  \ddots & \vdots \\
               0 &  0 &  \dots & U^{n-1}_{[1],[1]}}  =:U
\end{equation*}
Since $L^k$ is lower triangular with all ones on its diagonal and nonsingular,  the same hold for $\left( \prod_{k=1}^{n-1} L^k \right)^{-1}=:L$. Furthermore,  all diagonals of upper triangular $U$ are positive by assumption $A\in \mathcal{M}_n$. Therefore, all the leading principal minors of $A=LU$ are positive, which,    together with $a_{ij} \leq 0 $ from $\mathcal{M}_n$,  admit that $A$ is an M-matrix, and hence is a monotone matrix, i.e. $A^{-1}$ is non-negative,  see Exercise 8.3.P15 in \cite{horn2012matrix} and references therein. 
Consequently, $Ae  \preceq b \Rightarrow e \preceq A^{-1} b$.

To prove \eqref{Appendix:monotone}, it suffices to show that each entry of non-negative $A^{-1}=U^{-1}\prod_{k=1}^{n-1} L_k$ is not decreasing as some $-a_{ij}$, $i\neq j$,  increase. To this end, let us first study the monotony of  entries of $U^k$, then that of  $\prod_{k=1}^{n-1} L_k$ and $U^{-1}$.

Since $A$ has positive diagonals and non-positive off-diagonals and $U^k_{[1],[1]}$ is positive for all $k$ due to $A \in \mathcal{M}_n$, by exploiting the relation between $U^k$ and $U^{k-1}$, we can show iteratively that  
\begin{equation}\label{Appendix: U_k}
   U^k_{ii} >0, U^k_{ij}\leq 0, \ \forall i\in[1:n-k], \ j > i, k\in [1:n-1].  
\end{equation}
Assume that, for a fixed $k-1$,  each entry of $U^{k-1}$ does not increase as some  $-a_{ij}$ with $i \neq j$ increase.  Then, each entry of $ U^{k-1}_{[2:n],[2:n]}$ is non-increasing as well. Furthermore, each entry of   $U^{k-1}_{[2:n],[1]} U^{k-1}_{[1],[2:n]} $, which is non-negative due to \eqref{Appendix: U_k}, and each entry of $(U^{k-1}_{[1], [1]})^{-1}$, which is  positive by assumption,  are  non-decreasing.  Therefore,   each entry in $U^k$ is non-increasing. As each entry in $U^{0}=A$ is non-increasing with some increasing $-a_{ij}$, $i\neq j$, by induction,  we show that the same hold for   entries of $U^{k}$ for all $k\in [1:n-1]$. This together with \eqref{Appendix: U_k} exhibit   that  $\prod_{k=1}^{n-1} L^k$ is non-decreasing as some $-a_{ij}$ increase. Additionally, $\prod_{k=1}^{n-1} L^k$  is  non-negative due to \eqref{Appendix: U_k}.  Similarly, by showing that entries of  $(U_{[1:k],[1:k]})^{-1}$ do not decrease   if those of $(U_{[1:k-1],[1:k-1]})^{-1}$ do not with some increasing $-a_{ij}$,  one can  prove that each entry of  $U^{-1}$ is not decreasing with increasing $-a_{ij}$, $i\neq j$,  as well. Furthermore, $U^{-1}$ is also non-negative, since the upper triangular  $U$ is  M-matrix following \eqref{Appendix: U_k}.  Finally, we can conclude that each entry of $A^{-1}=U^{-1}\prod_{k=1}^{n-1} L^k$ is non-decreasing as some $-a_{ij}$, $i\neq j$,  increase.     
\end{proof}

\end{document}